\documentclass[11pt,titlepage]{article}
\usepackage{amsfonts,amssymb,amstext,amsthm,amsmath,amsbsy,amsopn,amsgen}

\newcommand{\ls}[1]
   {\dimen0=\fontdimen6\the\font
    \lineskip=#1\dimen0
    \advance\lineskip.5\fontdimen5\the\font
    \advance\lineskip-\dimen0
    \lineskiplimit=.9\lineskip
    \baselineskip=\lineskip
    \advance\baselineskip\dimen0
    \normallineskip\lineskip
    \normallineskiplimit\lineskiplimit
    \normalbaselineskip\baselineskip
    \ignorespaces
   }
\def\geq{\geqslant}
\def\leq{\leqslant}

\def\Blackboardfont{\mathbb}
\def\P{{\Blackboardfont P}}
\def\Q{{\Blackboardfont Q}}
\def\Z{{\Blackboardfont Z}}

\def\N{{\Blackboardfont N}}
\def\R{{\Blackboardfont R}}
\def\1{1\kern-0.30em 1}
\def\NN{I\kern-0.25em N}
\def\Im{{\rm Im}}

\def\sq{\hbox{\rlap{$\sqcap$}$\sqcup$}}
\def\qed{\ifmmode\sq\else{\unskip\nobreak\hfil
\penalty50\hskip1em\null\nobreak\hfil\sq
\parfillskip=0pt\finalhyphendemerits=0\endgraf}\fi}
\newcommand{\myline}{\newline\underline{\hskip\textwidth}}

\def\eref#1{(\ref{#1})}

\def\nopr#1{|#1|_{{\cal P}}}
\def\ov#1{\overline{#1}}
\def\uv#1{\underline{#1}}

\newcommand{\parag}{\medskip\noindent}
\newcommand{\eps}{\varepsilon}

\newcommand{\cqfd}{\hspace*{\fill}\rule{1.8mm}{1.8mm} \\ }
\newcommand{\intmin}[2]{\mbox{$\ominus\kern-2.1ex\displaystyle{\int_{#1}^{#2}}$}}

\DeclareMathAlphabet{\eus}{U}{eus}{m}{n}

\DeclareSymbolFont{cmrmn}{OT1}{cmr}{m}{n}
\DeclareMathAlphabet{\eurm}{U}{eur}{m}{n}

\def\eref#1{(\ref{#1})}

\def\<#1,#2>{\langle#1,#2\rangle}

\def\set#1#2{\{#1\mid\; #2\}}% ensemble des #1 t.q. #2 

\def\implies{\Rightarrow}

\def\cA{{\cal A}}
\def\cI{{\cal I}}

\def\cJ{{\cal J}}

\def\cB{{\cal B}}
\def\cF{{\cal F}}
\def\cE{{\cal E}}
\def\cG{{\cal G}}

\def\cC{{\cal C}}

\def\cL{{\cal L}}

\newtheorem{theo}{Theorem }[section]
\newtheorem{lemm}[theo]{Lemma }
\newtheorem{prop}[theo]{Proposition }
\newtheorem{defi}[theo]{Definition }

\newtheorem{coro}[theo]{Corollary}
\newtheorem{example}[theo]{Example }

\newenvironment{exam}{\begin{example}\rm}{\end{example}}
\newtheorem{remarkf}[theo]{Remarque }

\newtheorem{remark}[theo]{Remark }

\ls{1.15}

\title{Products of Irreducible Random Matrices in the (max,+)
Algebra\thanks{Supported by  
 the European Grant BRA-QMIPS of CEC
DG XIII.}}
\author{ Jean Mairesse\thanks{Research supported by the Direction des
Recherches 
Etudes et Techniques (DRET) under contract $n^{\circ}$ 91 815.}
(mairesse@sophia.inria.fr)  \\ 
INRIA Sophia-Antipolis \\ B.P.\ 93, 06902 Sophia Antipolis Cedex,
France \\ \\ 
{\it To appear in {\rm Advances in Applied Probability}, June 1997}}

\date{July 1995}

\begin{document}
\maketitle

\noindent
\begin{abstract}
We consider the recursive equation ``$x(n+1)=A(n)\otimes x(n)$''
where $x(n+1) $ and $x(n)$ are 
$\R^k$-valued vectors and $A(n)$ is an irreducible
random
matrix of size $k\times k$. The matrix-vector multiplication in the
(max,+) algebra is defined by $(A(n)\otimes x(n))_i= \max_j (A_{ij}(n)
+x_j(n))$. This type of equation can be used to
represent the evolution of Stochastic Event Graphs which include cyclic
Jackson Networks, some manufacturing models and 
models with general
blocking (such as Kanban). Let us assume that the sequence $\{A(n),\;n\in
\N\}$ is i.i.d or more 
generally stationary and ergodic. The main result of the paper states that the
system couples  
in finite time with a unique stationary regime if and only
if there exists a set of matrices $\cC$ such that $P\{A(0)\in \cC\}>0$
and the matrices $C\in \cC$ have a unique periodic regime.
\end{abstract}

\section{Introduction}
\label{se-intro}
Let us consider the following recursive equation:
\begin{equation}\label{eq-eq1}
\left\{
\begin{array}{ccc}
x_i(n+1)&=&\max_{ j} (A_{ij}(n) + x_j(n)) \\
x_i(0)  & = & (x_0)_i
\end{array}\right.
\end{equation} 
The sequences $\{A_{ij}(n)\}$ are given (exogenous data).
The process we want 
to
study is the sequence of  vectors $\{{\bf x}(n)=(x_1(n),\dots ,x_i(n),
\dots)'\}$. 
The vector $x_0$ is the initial condition. 

Because of the generality of Equation (\ref{eq-eq1}), it appears
in many different types of applications. In fact, this
equation appears in statistical mechanics in the study of crystal structures,
see Griffiths \cite{grif}.
It is also common under one form or another in economic and control
literature, see Yakovenko \& Kontorer \cite{YaKo}. In fact, it is the basic
Bellman equation of 
dynamic 
optimization in discrete time, for a finite state space. 
Recently, this kind of
equation has become very popular in the study of Discrete Events Dynamic
Systems (DEDS), see for example the recent textbooks of Baccelli, Cohen,
Olsder \& Quadrat \cite{BCOQ} and Glasserman \& Yao \cite{GlYa}. Let us
insist on 
the last point. 

\parag
A large class of computer or communication networks accepts a description as 
DEDS. Different approaches have been proposed to model DEDS. Petri Nets is
one of the most common formalism. 
More precisely, a sub-class of Petri Nets, Event Graphs, appears to be very
efficient in describing models with synchronization, blocking and/or fork-join
properties. We can mention Job-Shop models (see Cohen, Dubois, Quadrat \& Viot
\cite{CDQV85} or \cite{BCOQ}), cyclic
Jackson 
Networks (see Section \ref{sse:cjn}) or asymmetric exclusion
models as 
examples. On the other hand, Event 
Graphs cannot be used to model systems with routings.\\
We can describe the evolution of an Event Graph by the daters associated with
the 
transitions (nodes) of the graph.  It is well known, see for example
\cite{BCOQ},  that the
evolution of the daters of an Event Graph can be represented in
the form of Equation (\ref{eq-eq1}).\\
Generalized Semi-Markov Processes (GSMP) are another common formalism for
DEDS. It is 
shown in \cite{GlYa} that a GSMP with convex and homogeneous structural
properties 
admits a representation of the form (\ref{eq-eq1}).
In order to give a flavor of the modeling power
of Equation 
(\ref{eq-eq1}) and in order to motivate the practical interest of this work, 
we will present two different models in Section
\ref{se:tmm}. \\
\hspace*{1cm} $\bullet$ The first model appears in the modeling and the
analysis of parallel programs and architectures. It is a task graph with `and'
synchronizations also known as PERT Network.

\parag

\hspace*{1cm} $\bullet$ 
The second model is a closed cyclic Jackson
Network. We will consider both infinite and finite buffers with various
blocking modes.

\parag
It is very fruitful to use a matrix-vector notation for Equation
\eref{eq-eq1}. We define the following ``(max,+)'' notations:
\[
\forall x,y \in \R\cup\{-\infty\},\;\;x\oplus y=\max(x,y),\:x\otimes y=x+y\:.
\] 
We define also the $k\times k$ matrix $A(n)=\{A_{ij}(n),i,j=1,\dots, k\}$
and the column vector $x(n)=(x_1(n),\dots,x_k(n))'$.
With these notations,  the basic
Equation (\ref{eq-eq1}) takes a very simple and convenient form.
In fact, it can be rewritten as:
\begin{equation}
x(n+1) = A(n)\otimes x(n)\:.
\label{eq-eqmat}
\end{equation}
The matrix-vector product is defined in a natural way just by 
replacing  $+$ and
$\times$ by $\oplus$ and $\otimes$, i.e. $(A\otimes x)_{i}=\bigoplus_{j} A_{ij} \otimes x_j=\max_{j} (A_{ij} +
x_j)$. 

\parag

Historically, the first approach has been
to consider deterministic 
systems where $A_{ij}(n)\equiv A_{ij}$.

It is natural to consider a stochastic extension
where $\{A_{ij}(n)\}$ is a
sequence of random matrices.
As a consequence, here is an equivalent way of introducing our subject: it is
a counterpart of the classical theory of products of random matrices (see
Furstenberg \& Kesten \cite{FuKe} or Bougerol \& Lacroix~\cite{BoLa}) but in
another algebraic structure, the (max,+) algebra.

\parag
For systems described by Equation (\ref{eq-eqmat}), we will consider
two kinds of asymptotic results.

\parag

\hspace*{1cm} $\bullet$ First order limits, on ratios:
\begin{equation}
\lim_n \frac{\|x(n)\|_{\infty}}{n}\: , \;\;\lim_n \frac{x_i(n)}{n} \:.\label{eq-1ord}
\end{equation} 
\hspace*{1cm} $\bullet$ Second order limits, on differences:
\begin{equation}
\lim_n \: x_i(n+1)-x_i(n),\;\forall i,\;\;
\lim_n \: x_j(n)-x_i(n),\; \forall i\neq j\:.\label{eq-2ord}
\end{equation} 

A first order limit is a cycle time or equivalently the inverse of a
throughput.
Second order limits
are related to waiting and idle 
times, queue length and frequency of occupation.
More insights on the relations between these limits and quantities of interest
for the system will be provided in Section \ref{se:tmm}. Our 
goal is to find 
stationary regimes for second order limits. Multiple
stationary regimes will mean multiple possible regimes for waiting
times or queue lengths, depending on the initial condition.

\parag
Among the systems modeled by Equation (\ref{eq-eqmat}), we can distinguish
two classes: the open (or non-autonomous) systems and the closed (or
autonomous) ones. Open systems have been exhaustively treated by  
Baccelli~\cite{bacc92}~\cite{BCOQ} (for both first and second order
limits).  
Problems of existence and uniqueness of first order limits for closed systems
have been 
solved by Cohen~\cite{cohe} (see also \cite{bacc92}). These results are
recalled in \S \ref{sse:rfbacc}. 
This paper deals with the open question of existence and uniqueness of second
order limits  for
closed systems. These problems were considered
in several earlier
papers (Resing, de Vries, Hooghiemstra, Keane \& Olsder \cite{Ral90} and
\cite{Oal90}, Baccelli \cite{bacc92}) but only sufficient 
conditions of
uniqueness were known. 
The approach we use is new
and exploits completely the common hidden algebraic structure of the
different models we consider. It enables us to obtain necessary and
sufficient conditions for stability (in some cases) together with simple proofs.

\parag

The conditions we give
are based on the structure of
deterministic matrices chosen in the support of the random matrix
$A(0)$. One of the main results states that the
system couples  
in finite time with a unique stationary regime if and only
if there exists a set of matrices $\cC$ such that $P\{A(0)\in \cC\}>0$
and the matrices $C\in \cC$ have a unique periodic regime.
The proof makes use of Borovkov's theory of renovating events, see Borovkov \&
Foss \cite{BoFo92}~\cite{BoFo94}.
This theory 
appears to be much more tractable than classical Harris regeneration
due to the specific form of our recursive equations. 

\parag

The paper is organized as follows. We introduce two models
in Section
\ref{se:tmm}, cyclic Jackson
Networks and task graphs with random precedences.  Sections \ref{se:m+a},
\ref{se:dst} and 
\ref{se:bret} are presenting the tools that we are using in the paper. They 
can be skipped by people knowing the subject. Section \ref{se:m+a} is devoted
to the (max,+) algebra, Section \ref{se:dst} to the spectral 
theory in this algebra and Section \ref{se:bret} to Borovkov's theory of
renovating 
events. Section \ref{se:prr} presents the main results. In \ref{sse:rfbacc},
we recall some results from \cite{bacc92} and \cite{cohe}.
In \ref{sse:prr}, we state
some 
preliminary results. In \ref{sse:sdm} and \ref{sse:sgm}, we give 
sufficient 
conditions for the stability of discrete and general models respectively.
In Section \ref{se:ct}, we establish the converses of the results of
the
previous section. In Section \ref{se:wi}, we weaken the assumptions under
which some of our results apply.
Finally, for convenience,
some of the proofs are given in 
Appendix.

\section{Two Motivating Models}
\label{se:tmm}

\subsection{Task graphs}
\label{sse:tg}
We consider a parallel program executed on several identical 
processors. We model it
by its precedence graph $\tau$. If we consider a system of $k$ processors, 
the graph $\tau$ has a set of nodes which is $ k \times \N$. The node  $(i,n)$
represents the $n$-th task to be executed at processor $i$. The arcs
between nodes represent 
the synchronization constraints. There
is an arc between
the node $(i,n)$ and the node $(j,m)$ (notation~: $(i,n)\rightarrow (j,m)$) 
if the $n$-th task at processor $i$ has to be
completed in order for the $m$-th task at processor $j$ to be enabled.
The execution of a task begins as soon as all the tasks 
of its incoming arcs are
completed. 
Each task has a
duration which may depend on the processor.

\parag
Let us consider a task graph with synchronizations only between consecutive
levels $n$ (i.e.
nodes $(1,n),\dots,(k,n)$) and $n+1$. 
We assume that the synchronizations depend on $n$.
We denote by ${\cal L}(i,n)$ the set of nodes $j$
such 
that $(i,n)\rightarrow (j,n+1)$.
We suppose that for all $i$, there exists a probability law $P^i$ 
on the
subsets of $(1,\dots, k)$ such that ${\cal L}(i,n)=(j_1,\dots,j_p)$ with
probability $P^i\{(j_1,\dots,j_p)\}$. 
We denote by $x_i(n)$ the
date of completion of task $n$ at processor $i$, and by $A_{ji}(n)$ the
duration 
of the synchronization constraint between nodes $(i,n)$ and $(j,n+1)$ (it
may include a transmission time as well as the execution time at processor
$j$). 
We adopt the convention that 
$A_{ji}(n)=-\infty$ if  $j \not\in {\cal L}(i,n)$. 
It is easy to check that such a model,
we could call it a task graph with random precedences, verifies Equation
(\ref{eq-eqmat}). \\

A Queuing Network model studied by Baccelli \& Liu \cite{BaLi92a}
corresponds
to this 
model. The task resource models studied in \cite{BrVi} or \cite{GaMa95} 
also have this kind of structure.

\subsection{Cyclic Jackson network}
\label{sse:cjn}
We consider a closed Jackson Network. The study of such closed networks can be
traced
back to Gordon and Newell, \cite{GoNe}. In their original model,
there is 
a given number of
indistinguishable customers. The routing of  the customers leaving a given
queue 
is provided by a sequence of i.i.d. Bernouilli random variables. All the
service times are exponential.
They prove the existence of an explicit product form for the
unique stationary distribution. 

\parag
A natural generalization of the basic model is to consider
i.i.d. (resp. stationary and ergodic) sequences of service
times with general distributions, i.e. to replace $./M/1/\infty$ servers by
$./GI/1/\infty$ (resp. $./G/1/\infty$) servers. Finding the minimal assumptions
leading to a unique 
stationary regime for this generalized closed Jackson Network is still an open
problem. 

\begin{figure}[htb]
\centerline{\input{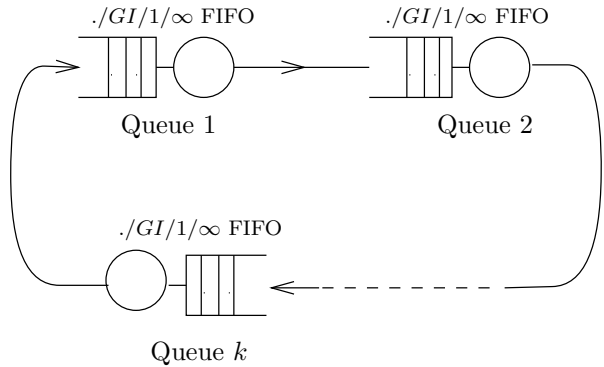}}
\caption{\sf A Cyclic Jackson Network consisting of $k$ queues.
\myline}
\label{fi:jak}
\end{figure}

\parag 
We consider a restriction of the previous model. There are $k$ queues and all
customers have the same cyclic route $(1,2,\dots,k,1)$,  see Figure
\ref{fi:jak}. 
We will
denote this model by CJN for Cyclic Jackson Network, following the terminology
of \cite{KaMa92}.

In the following, the numbering of
queues has to be understood modulo $[k]$, for example queue $(k+2)$ is queue 2.
We denote by $\{\sigma_j(n),n\in \N\}$, the sequence of service times at queue
$j$. 
%This sequence is i.i.d. or stationary and ergodic.
%or more generally stationary and ergodic. 
%We suppose also that the service times at the different queues are
%independent. 
Instead of describing the system by the workload or the queue length process,
as is usually done, we propose to study this model by introducing the
following variables. With 
each queue $j$, we associate a dater $\{x_j(n),n\in \N\}$. The variable
$x_j(n)$ represents the date of completion of the $n-$th service at queue
$j$. All variables of interest for the network can be derived from
these 
daters and from the  sequences of service times.
More 
precisely, we have:
\begin{itemize}
\item Asymptotic throughput at queue $j$:
\[
\gamma_j=\lim_{n\longrightarrow +\infty} \frac{n}{x_j(n)}\:.
\]
\item Idle time of queue $j$ before the arrival of the $n$-th customer to visit
queue $j$. 
\[
I_j(n)=x_j(n)-\sigma_j(n)-x_j(n-1)\:.
\]
\item Workload at queue $j$ at the instant of the arrival of the $n$-th
customer to visit queue $j$. This customer comes from queue $j-1$. We suppose
that it was the $n'$-th customer to visit queue $j-1$.
\[
W_j(n)=x_j(n)-\sigma_j(n)-x_{j-1}(n')\:.
\]
\end{itemize}
The variables ($\gamma_i$) which are obtained as ratios of daters will be
called first order 
variables. The ones ($I_j,W_j$) which are obtained as differences of daters
will be called 
second order variables. We are concerned with the problem of deriving
necessary and sufficient conditions under which there is a unique stationary
regime for both first and second order variables. In such a case, we say
that our model is {\it stable}.

\parag
Suppose for the moment that there are exactly
$k$ customers. We suppose also that there is initially one customer in each
queue.  These assumptions together with the FIFO service discipline at each
queue yields the following property. The $n$-th customer to visit queue $j$
will be, at the next step of its route,  the $(n+1)$-th customer to visit queue
$j+1$. 
As a consequence, we have
\[
\left\{
\begin{array}{ccc}
x_1(n+1) & = & \max (x_1(n),x_k(n)) + \sigma_1(n)  \\
x_2(n+1) & = & \max (x_2(n),x_1(n)) + \sigma_2(n)  \\
 & \cdots & \\
x_j(n+1) & = & \max (x_j(n), x_{j-1}(n)) + \sigma_j(n) \\
 & \cdots & 
\end{array}
\right. \:.
\]
Using the (max,+) notation, this can be
rewritten as:
\begin{equation}
\label{eq-cjn}
x(n+1)=A(n)\otimes x(n),\; {\rm where }\;A(n)=\left(
\begin{array}{ccccc}
\sigma_1(n) & -\infty & \cdots &-\infty &\sigma_1(n) \\
\sigma_2(n) & \sigma_2(n) & \ddots &  &-\infty \\
-\infty & \ddots & \ddots & -\infty &\vdots  \\
\vdots & \ddots &\ddots & \ddots & -\infty  \\
-\infty & \cdots &-\infty & \sigma_k(n) & \sigma_k(n)
\end{array}
\right) \:.
\end{equation}
The initial condition is $x(0)\geq 0$, where  $x_i(0)$ is the remaining
service time of the customer being served at queue $i$ at time 0.

\parag
When the service times are
deterministic, it is possible to obtain many asymptotic behaviors, depending
on the initial condition $x(0)$. In fact, initial delays between customers
might never vanish. Therefore, it is possible to
have several 
stable regimes 
for second order quantities ($I_j,W_j,\dots$) including periodic ones. For
stochastic systems, when the service times are random variables, it is still
possible to have several stationary regimes if the system is not ``stochastic
enough''.  As an application of the results
presented in this paper, we obtain the necessary and sufficient conditions
for the existence of a unique stationary regime for this CJN.
This model will be used as an illustration of the
results throughout the paper (Examples \ref{ex:jak1}, \ref{ex:jak2},
\ref{ex:jak3}, \ref{ex:jak4}).

\parag
When there are less than $k$ customers in the network,
the system can be represented in the same way as previously. The only
difference is that the structure of matrices $\{A(n)\}$ is more complicated.
When there are more than $k$ customers, the trick consists in splitting
queues. Each
queue which has originally $(p>1)$ customers in its buffer is transformed into
$p$ queues with one customer per buffer. This
is done by creating $p-1$ fictive queues with service times identically
equal to zero. By doing this, one gets back to the previous case. The
main difference is that we have represented our model by a (max,+) linear
system of dimension greater than the original number of queues. For more
details on these transformations, see \cite{mair95e}. 

\parag

We can 
also model CJN with 
finite buffers (the $./G/1/L$ case).
Finite buffers imply the blocking of some customers. There are different
possible types of blocking.
\begin{enumerate}
\item Blocking before service. The service begins
at queue $i$ only when the buffer at queue $(i+1)$ is not full.
\item Blocking after service (of communication type). Service at queue $i$
begins as soon as a customer is available. After completion of the service,
if the buffer of queue $(i+1)$ is full, the customer starts another service at
 queue $i$.
\item Blocking after service (of manufacturing type). Service at queue $i$
begins as soon as a customer is available. After completion of the service,
the customer has to wait in queue $i$ if the buffer of queue $(i+1)$ is full.
It prevents another customer from being served at queue $i$.
\item Blocking after service (of Kanban type). The mechanism is the same as
previously but there exists a finite intermediate buffer between queue $i$ and
$(i+1)$. A 
customer completing its service at queue $i$ enters this intermediate buffer
if 
the buffer of queue $(i+1)$ is full and the intermediate buffer is not. 
It enables to serve a new 
customer at queue $i$.
\end{enumerate}
Excepting the blocking after service of communication type, all these types of
blocking can be considered. We can even consider different types of blocking
for the different queues of the network. In all cases, we obtain 
a (max,+) 
linear representation for the network.

\parag
In the case of a CJN with i.i.d. general
service 
times (./GI/1 servers), the classical method for studying the
network is to
consider the Markov chain formed by queue lengths and
remaining service times and to apply Harris regeneration techniques. 
This method was first introduced for closed
acyclic Jackson Networks by Borovkov \cite{boro86}, \cite{boro88}. For closed
Cyclic Jackson Networks, it is used by
Bambos \cite{bamb} and 
Kaspi $\&$ Mandelbaum \cite{KaMa92}~\cite{KaMa94}.
The method provides sufficient conditions of stability.
Our approach is completely different and provides
necessary and sufficient
conditions of
stability for CJN.

\parag

For a much more complete description of the systems modeled by 
Equation (\ref{eq-eqmat}), the
reader is referred to 
the textbook of Baccelli, Cohen, Olsder and Quadrat \cite{BCOQ}.

\section{(max,+) Algebra}
\label{se:m+a}
\begin{defi}[(max,+) algebra]
We consider the semiring $(\R\cup\{-\infty\},\oplus,\otimes)$.
The law $\oplus$ is ``max'' and $\otimes$ is the usual addition.
We set 
$\eps=-\infty$ and $e=0$.
The element $\eps$ is neutral for the operation $\oplus$ and absorbing for
$\otimes$. 
The element $e$ is neutral for $\otimes$. The law $\oplus$ is idempotent, i.e.
$a\oplus a=a$. $(\R\cup\{\eps\},\oplus,\otimes)$ is an idempotent semiring,
called 
a dioid. It is usually referred to as the
(max,+) algebra.
We shall denote it by $\R_{max}$. 
\end{defi}

We define the product spaces $\R_{max}^k,\:\R_{max}^{k\times k}$. We
define the product of a vector by a scalar: $a\in \R_{max},u \in
\R_{max}^k,\;(a\otimes u)_i = a\otimes u_i$. \\
Matrix product is defined in a natural way, replacing + and $\times$ by
$\oplus$ and $\otimes$ respectively.
Let $A,B \in \R_{max}^{k \times k}$,
\[
(A\otimes B)_{ij}=\max_{l} (A_{il} + B_{lj})=\bigoplus_{l} A_{il} \otimes
B_{lj}\:. 
\]
Matrix-vector product is defined in a similar way, see Section \ref{se-intro}.
In the rest of the paper, the notations ``+,$\times$'' will stand for
the usual addition and multiplication.
Nevertheless, we will write $ab$ for $a\otimes b$ whenever there is no
possible confusion. For example, for $A\in \R_{max}^{k\times k},
\;A^d=A^{\otimes d}=\underbrace{A\otimes \cdots \otimes A}_{d
\;\mbox{factors}}$.

\parag

Let us recall some definitions adapted to the $\R_{max}$ algebra.

\begin{defi}
The graph of a square matrix $A$ is a directed graph having a number of
nodes equal to the size of $A$. This graph contains an arc from $i$ to $j$ iff
$A_{ji} \neq \eps$. The valuation of this arc is $A_{ji}$. 
\label{de:comgr}
\end {defi}

\begin{defi}
A square matrix $A$ is
irreducible if:  
$\forall i,j \;\;\exists n \geq 0\;\;|\;\;(A^n)_{ij} > \eps$ (or equivalently
if its graph is strongly connected). 
\end {defi}

\begin{defi}
\label{ape}
A square matrix $A$ is aperiodic if:  $
\exists N, \;\forall n\geq N,\;\forall i,j,\;\;\; (A^n)_{ij}>\eps$.
\end {defi}

\begin{defi}
Let $(\Omega, {\cal F}, P)$ be a
probability space. 
A stochastic matrix $\{A(\omega),\;\omega \in \Omega\}$ has a fixed structure
if  
$ P(A_{ij}=\eps)=1$ or $ 
P(A_{ij}=\eps)=0 ,\;\forall i,j$.
\label{struc}
\end {defi}

\begin{defi}[$\P\R^k$]
The projective space $\P\R^k$ is defined as the quotient of $\R^k$
by the parallelism relation: 
\[
u,v\in \R^k\;\;\;\;u\simeq v \Longleftrightarrow \exists a \in
\R 
\;{\rm such\;that}\;u=a\otimes v\:.
\]
Let $\pi$ be the canonical projection of $\R^k$ into $\P\R^k$.
\label{de:pi}
\end{defi}
For example $(e,-1)'$ and $(2,1)'=(e+2,-1+2)'$ are in the same parallelism
class, i.e. are two 
representatives of the same element of $\P\R^k$. 
We define in the same way $\P\R^k_{max}$, $\P\R^{k\times k}_{max}$ and
$\P\R^{k\times 
k}$. 
We use the same notation $\pi$ for the different canonical projections.
We define a norm and a distance on $\P\R^k$ which we call 
the projective norm and 
distance. 
\begin{defi}
Let $x\in \P\R^k$ and $u\in\R^k $ be a representative of $x$, i.e. $\pi(u)=x$.
We define:
\[
\nopr{x}=\max_i{u_i} - \min_i{u_i}\:.
\]
Let $x,y \in \P\R^k$ and $u,v \in \R^k$ be two representatives of $x$ and $y$
respectively.
We define:
\[
d(x,y)=d(u,v)=\nopr{x-y}=\bigoplus_{i}(u_i-v_i)\:\otimes \: \bigoplus_{i}(v_i-u_i)\:.
\]
We write either
$d(x,y)$ or $d(u,v)$ with some abuse of notation.
\label{dist}
\end{defi}

The space $(\P\R^k,\nopr{.})$ is an Euclidean space. 
In particular, it is complete.
The norm $\nopr{.}$ corresponds to
the ${\cal L}_{\infty}$ norm\footnote{It is worth
mentioning that $d(.,.)$ is the $\R_{max}$ analogue of a 
distance used in classical Perron-Frobenius 
theory, which is called the Hilbert's projective metric and is defined by ``$\delta(u,v)=\ln \left(
\inf\{\mu/\lambda\:|\:\lambda u
\leq v \leq 
\mu u\}\right)$''.} on the projective
space $\P\R^k$.  

\begin{prop}
Let $A \in \R^{k\times k}_{max}$ be an irreducible matrix. Let $u,v$ be two
vectors of $\R^k$. We have: 
\[
d(Au,Av) \leq d(u,v)\:.
\]
\label{decd}
\end{prop}

\begin {proof}
By definition, we have:
\[
d(Au,Av)=\bigoplus_{i}((Au)_i-(Av)_i)\:\otimes \:
\bigoplus_{i}((Av)_i-(Au)_i) \:.
\]
We define $j(i)$ such that $(Au)_i=\bigoplus_{j} A_{ij}\otimes
u_j=A_{ij(i)}\otimes u_{j(i)}$. Note that $j(i)$ depends on $A$ and $u$. We
have: 
\begin{eqnarray*}
\bigoplus_{i}((Au)_i-(Av)_i)&=&\bigoplus_{i}\left(\:(\bigoplus_{j}
A_{ij}\otimes 
u_j)\:-\:(\bigoplus_{j} A_{ij}\otimes
v_j)\:\right) \\
    & =& \bigoplus_{i}\left(\:( A_{ij(i)}\otimes
u_{j(i)})\:-\:(\bigoplus_{j} A_{ij}\otimes v_{j})\:\right) \\
    & \leq & \bigoplus_{i} (A_{ij(i)}\otimes u_{j(i)} - A_{ij(i)}\otimes
v_{j(i)}) \\
  &=& \bigoplus_{i} u_{j(i)}- v_{j(i)} \leq  \bigoplus_{i} u_i -v_i 
\end{eqnarray*}
We obtain
$
d(Au,Av)    \leq \bigoplus_{i}( u_i -v_i )\:\otimes \: \bigoplus_{i}( v_i
-u_i )$ i.e. $ d(Au,Av)    \leq d(u,v)$. 
\end {proof}

There is no simple criterion to get a strict inequality. This monotonicity 
has to be 
interpreted as a synchronization property.

\begin{defi}
We consider $A\in \R^{k\times k}_{max}$. We set 
\[
{\tt D}(A)=\sup_{u,v\in \R^k} d(Au,Av)\:.
\]
We call ${\tt D}(A)$ the projective diameter of $A$.
\label{D}
\end{defi}
It is easy to prove that ${\tt D}(A)$ is finite if and only if
$\forall i,j,\; A_{ij} >\eps$. 
A matrix $A$ can be considered as a ``linear'' (in the (max,+) sense)
operator
from
$\P\R^k$ into $\P\R^k$. It is a bounded operator if the (decreasing) sequence
${\tt D}(A^n)$ 
has a finite limit, i.e. if $A$ is aperiodic (Def. \ref{ape}).

\section{Deterministic Spectral Theory}
\label{se:dst}
We recall some results of the deterministic spectral theory in the
$\R_{max}$ 
algebra. Theorem \ref{lyap} is due to
Cuninghame-Green~\cite{cuni62}. Versions of Theorem
\ref{vectprop} were proved in~\cite{roma}~\cite{cuni79} and
\cite{GoMi77}. Under 
the form proposed here, the result is  
from \cite{CDQV85}.
Theorem \ref{thcyc} is due to Cohen, Dubois, Quadrat and
Viot~\cite{CDQV83} and \cite{CDQV85}. A complete
treatment of the spectral theory can be found in \cite{BCOQ}.

\parag
We want to find non trivial solutions to the equation:
\[   A\otimes x= \lambda\otimes x \:, \]
where $A \in \R^{k \times k}$ is an irreducible matrix, $x$ is a column vector
(the 
``eigenvector'') and $\lambda$ is a real constant (the ``eigenvalue''). We
define also periodic regimes for the eigenvalue 
problem. A periodic regime of period
$d$  is a set of vectors $\{x_1,\dots ,x_d  \}$ of
$\R^k$ verifying $Ax_i=\lambda x_{i+1},\;i=1,\dots,d-1$ and $Ax_d=\lambda
x_1$. 

\begin{defi}
For each path $\zeta=\{t_1,t_2,\cdots ,t_j,t_{j+1}=t_1\}$, we define its
average 
weight by:
\begin{displaymath}
p(\zeta)=a_{t_1t_j} \otimes \cdots \otimes
a_{t_{3}t_{2}} \otimes a_{t_{2}t_{1}}/j\:,
\end{displaymath}
(here the division is the conventional one).
\end{defi}

\begin{theo}
There is a unique (non $\eps$) eigenvalue, $\lambda$. It satisfies the relation
\[ \lambda=\max_{\zeta} \; p(\zeta)\:,  \]
where $\zeta$ covers all the circuits of (the graph of) $A$.
We call also $\lambda$ the Lyapunov exponent or the cycle time of $A$.
\label{lyap}
\end{theo}

There might be several eigenvectors. A linear combination (in $\R_{max}$) of
eigenvectors is an eigenvector. An eigenvector has all its coordinates different from
$\eps$ (due to the irreducibility assumption). 

\begin{defi}
For a matrix $A$, we define:
\begin{description}
\item [Critical circuit] A circuit $\zeta$ of $A$ is said to be critical if its
average weight is maximal, i.e. if $p(\zeta)=\lambda$.
\item[Critical graph] It consists of the nodes and arcs of $A$ belonging to the
critical circuit(s).
\end{description} 
For a general graph, we define :
\begin{description} 
\item [Cyclicity] The cyclicity of a strongly connected graph 
is the greatest common divisor of the lengths of all the circuits. The
cyclicity of 
a connected graph is the least common multiple of the cyclicities of its
maximal strongly connected 
subgraphs (s.c.s.).
\end{description}
\label{3def}
\end{defi}

We normalize a matrix by subtracting (in the
conventional algebra) the eigenvalue to all the coordinates. 
The eigenvalue of a
normalized matrix is $e$.
For a normalized matrix $A$ of size $k$, we define:
\[
A^{+} =  A \oplus A^2 \oplus \cdots \oplus A^k\:.
\]
We check that $A^{+} \oplus A^{k+1}=A^{+}$.
\begin{theo}
Let $A$ be a normalized matrix. 
\begin{enumerate}
\item[a.] Critical columns $A^{+}_{.i}$, $i$ belonging to the critical graph,
are eigenvectors.
\item[b.] For $i,j$ belonging to the critical graph, $\pi(A^{+}_{.i})$ and
$\pi(A^{+}_{.j})$ are different iff  
they belong to two different s.c.s. of the critical graph.
\item[c.] Every eigenvector of $A$ writes as a linear
combination (in $\R_{max}$) of critical columns
$A^{+}_{.i}$.
\end{enumerate}
\label{vectprop}
\end{theo}

In $\R_{max}$, every irreducible matrix is cyclic in the sense of the following
theorem.
\begin{theo}
For an irreducible matrix $A$ of size $k$ and whose eigenvalue is $\lambda$,
there 
exists integers $d$ and $M$ such that:
\begin{equation}
\forall m \geq M,\;\;\;A^{m+d}=\lambda^{\otimes d}\otimes A^m\:,
\label{eq-cyc}
\end{equation}
furthermore the smallest $d$ verifying the property is equal to the cyclicity
of the 
critical graph of $A$. We call it the {\bf cyclicity} of $A$. \\
\label{thcyc}
\end{theo}

A cyclicity greater than one will provide periodic regimes of period greater than
one for the eigenvalue problem.

\begin{prop}
An irreducible matrix has a unique eigenvector and no periodic regimes of period greater than
one for the eigenvalue problem, if and only if its critical graph has a unique s.c.s. and its cyclicity is one. Such a matrix will be called a
{\bf scs1-cyc1} matrix.
\label{pr:scscyc}\index{scs1-cyc1}
\end{prop}

The proof follows from Theorems \ref{vectprop} and \ref{thcyc}.

\begin{defi}[rank]
\label{de:rank1}
By analogy with 
classical linear algebra, we define 
the ``rank'' of a matrix $A$ as the number of additively
independent columns (resp. lines) of $A$. More precisely, 
let $A_{.i}$  denote the $i$-th column  of $A$.
Matrix $A$ is of rank $r$ if there exists $\cJ \subset \{1,\dots, k\}$
such that $|\cJ|=r$ and $\forall i\neq j \in \cJ, \; \pi(A_{.i})\neq \pi(A_{.j})$ and
$\forall i \not\in \cJ,~\exists \alpha_j, j\in \cJ$, such that
\[
\pi(A_{.i})=\pi[\bigoplus_{j\in \cJ}  \alpha_j \otimes A_{.j} ]\:.
\]
\end{defi}

Let $A$ be a {\bf rank 1} matrix. Then $A$ is a {\bf scs1-cyc1} matrix and
verifies $A^2=\lambda \otimes A$ ($\lambda$ is the eigenvalue of $A$).
The other way round, let $A$ be a {\bf scs1-cyc1} matrix and $M$ 
be defined as in Equation \eref{eq-cyc}. One can
check that $A^M$ is a matrix of {\bf rank 1}.

\begin{exam}[Cyclic Jackson Network 1] 
\label{ex:jak1}
Let us consider a basic Cyclic Jackson Network as presented in Section
\ref{sse:cjn}. We suppose that the service times are deterministic, i.e
$\sigma_j(n) \equiv \sigma_j$. We suppose also that the number of customers,
$k$, is 
equal to the number of queues. Then we can consider the (max,+) matrix
associated with the network, see Equation (\ref{eq-cjn}). 
The graph associated with this matrix is constituted by the
circuit $(1,2,\dots, k,1)$ and the recycling loops $(1,1)$ to $(k,k)$. 
Let us define $I=\set{ i }{ \sigma_i=\max_j \sigma_j }$. There are two possible
cases.
\begin{itemize}
\item If the cardinal $| I |< k$, then the critical graph of the matrix consists of the nodes $i\in I$ and the arcs $(i,i), i\in I$. It implies
that the matrix is scs$|I|$-cyc1.
\item If $| I |= k$ then the graph and the critical graph of the matrix
coincide. It implies that the matrix is scs1-cyc1.
\end{itemize}
We conclude that the matrix is scs1-cyc1 if and only if $|I|=1$ or $k$. 
\end{exam}

\section{Borovkov's Renovating Events Theory}
\label{se:bret}
Borovkov's theory deals with the problem of regeneration in so-called
``Stochastic Recursive 
Sequences''. For a complete treatment, the reader is referred to Borovkov
\cite{boro84}, Borovkov \& Foss \cite{BoFo92,BoFo94}
or Brandt, Franken \& Lisek \cite{BFLi}. Let $(\Omega,{\cal
F},P)$ be a probability space. Let $\theta$ be a measurable map from
$(\Omega,{\cal F})$ into itself such that $P$ is $\theta$-invariant and
$\theta$-ergodic. Let $(E,\cE)$ and $(G,\cG)$ be two Polish spaces 
(complete, separable metric spaces) equipped with their respective Borel 
$\sigma$-algebra. 

\begin{defi}
We call Stochastic Recursive
Sequence (SRS), a sequence $\{x(n)\}$ of $E$-valued random variables
defined by  
\[ 
x(n+1)=f(x(n),a(n)),\;\; n\geq0, \;x(0)=x_0\:,
\]
where  $\{a(n)\}$ is an exogenous sequence of $G$-valued random variables,
stationary with respect to
the shift $\theta$. The function $f$ is a measurable function from
$E\times G$ into 
$E$. The vector $x_0\in E$ is the initial condition. In order to stress the
value of the initial condition, 
we will sometimes denote the SRS by $\{x(n,x_0)\}$.
We talk of an i.i.d. SRS when the sequence $a(n)$ is i.i.d. (an i.i.d.
SRS is a Markov 
Chain and the converse is true).
\label{srs}
\end{defi}

\begin{defi}
We consider a SRS $\{x(n)\}$. We denote by ${\cal F}_{l}$ the
$\sigma$-algebra ${\cal F}_{l}=\sigma 
\{a(n),$ $ n \in \{ -\infty,\dots,l-1\}\}$. 
The sequence of events $\{{\cal A}(n) \in
{\cal F}_{n+m},\;n\in \N\}$ is said to be a renovating sequence of length $m$
and of associated function  
$\phi:G^m \rightarrow E$ if:
\[
\exists n_0, \;\forall n \geq n_0,\;
x(n+m)=\phi\left(a(n),a(n+1),\cdots,a(n+m-1)\right)\;\;{\rm on}\;\; 
{\cal A}(n)\:.
\]
A sequence $\left\{{\cal A}(n),n\in \N\right\}$
of renovating events of same length and associated function is said to be
stationary if 
${\cal A}(n)= {\cal A}(0)\circ\theta^n=\theta^{-n} {\cal A}(0)$.
\end{defi}

We need the following notions of convergence:

\begin{defi}
We say that there is coupling convergence in finite time (or, merely,
coupling) of a sequence $\{X_n\}$  	
to a stationary sequence $\{Y \circ \theta^n\}$ if
\[
P(X_{n+l}=Y\circ \theta^{n+l},\;\forall l\geq
0)\stackrel{n\rightarrow
+\infty}{\longrightarrow}1\:.
\]
\end{defi}
It is easy to show that this notion of coupling convergence implies total
variation convergence 
($X_n \rightarrow Y$ in total variation if $ \sup_{A\in {\cal F}}\:|P(X_n \in
A)-P(Y \in 
A)|\:\stackrel{n\rightarrow +\infty}{\longrightarrow}0$).       
                          
\begin{defi}
We say that there is strong coupling convergence in finite time (or, merely,
strong coupling) of a sequence $\{X_n\}$
to a stationary sequence $\{Y\circ\theta^n\}$ if:
\[
\nu ={\rm min}\left\{n\geq 0 \;|\; X_{n+l}\circ\theta^{-(n+l)}=Y,\;\forall l
\geq 0\right\}\;\; {\rm is \;\: a.s}\;\:{\rm finite}\:.
\]
\label{scoup}
\end{defi}
%
%Strong coupling can be defined in another way:
%$\mu$ a.s. finite, with:
%\[
%\mu=\sup_{n \geq 0} \mu_n \; {\rm o\grave{u}} \; \mu_n=\min \left\{n\geq 0 \;|\;
%X_n\circ\theta^{n}=Y\circ \theta^{n+n}\right\}\:.
%\]
% 
{\bf Remark } Strong coupling implies coupling but the converse is not true.

\begin{theo}[Borovkov's renovating events]
\label{thbo1}
%Let $(\Omega,{\cal F},P)$ be a probability space. Let $\{a(n)\}$
%be a sequence of $\theta$-stationary random variables, $\R^J$-valued. 
We consider a SRS
$\{x(n)\}$ defined by:
\[
x(n+1)=f(x(n),a(n)),\;\; n\geq0,\;x(0)=x_0\:.
\]
If the random process $\{x(n),n\in \N\}$ admits a stationary sequence of
renovating events  
$\left\{{\cal A}(n)\right\}$ such that $P({\cal
A}(0))>0$, then there exists a finite random variable $Z$ such that:
\[
Z\circ \theta=f(Z,a(0))\:,\;\;
\]
and $x(n)$ converges with strong coupling in finite time to $ Z\circ
\theta^n$. 
\end{theo}

%\begin{theo}[converse of Th. \ref{thbo2}]
%\label{thborec}
%Conditions of Theorem \ref{thbo} (i.e. there exists a stationary sequence of renovating events $\{ {\cal A}_n \}$
%with $P( {\cal A}_0)>0$) are  necessary and sufficient for strong coupling convergence.
%\end{theo}
%
%\parag

In the previous theorem, we have considered a SRS defined with a unique
initial condition, $x_0$. In the rest of 
the paper, we will be interested in having results that hold uniformly over
the initial 
conditions. We will then use the 
following generalization of Borovkov's theorem.

\begin{theo}
\label{thbo2}
We consider a subset $V$ of $E$ ($V=E$ is in particular possible).
We suppose that  there exists a stationary sequence of 
events $\{{\cal A}(n)\}$ verifying $P({\cal A}(0))>0$ and which is renovating
for the SRS $\{x(n,x_0)\}$,  
$\forall x_0 \in V$. 
Then, for all (possibly random) initial condition 
$x(0)$ such that $P(x(0)\in V)=1$, the sequence $\{x(n)\}$ 
converges with strong coupling to a unique stationary 
regime.
\end{theo}

\begin{theo}[converse of Th. \ref{thbo1} and \ref{thbo2}]
\label{thbo2c}
The conditions of Theorem \ref{thbo1} are  necessary and sufficient for strong
coupling 
convergence. Let $V$ be a compact subset of $E$. The conditions of
Theorem \ref{thbo2} are  necessary and sufficient for strong coupling
convergence 
uniformly
over initial conditions in $V$.
\end{theo}

Next theorem was proved
by Anantharam and Konstantopoulos in \cite{AnKo}.

\begin{theo}
\label{anko}
Let $(\Omega,\cF, P)$ be a probability space. We assume that $(\Omega,\cF)$ is
a Polish space equipped with its Borel $\sigma$-algebra.
We consider a SRS ``$x(n+1)=f(x(n),a(n))$'' defined on $E$.
Suppose that,
for some $x_0\in E$, the sequence $\{x(n,x_0)\}$
is tight\footnote{Tightness on $E$ means that for any $\eta >0$, there is a compact $K$ of $E$ such that $P\{x(n,x_0) \in K \}>1-\eta$, for all $n$.} on $E$.
Then there is a stationary distribution for the SRS.
\end{theo}
The stationary
distribution is defined on $\Omega\times E$ with an $\Omega$ marginal 
equal to $P$. It
provides only a {\it weak
stationary regime} ({\it wsr}) for the SRS, 
see \cite{AnKo} or \cite{BFLi} for details. 
All we need to know about {\it wsr} is that
stationary regimes are {\it wsr}. Hence, the uniqueness of
stationary regimes implies the uniqueness of {\it wsr}. 

\parag
It is proved in \cite{BoFo92}, that for an i.i.d. SRS (i.e. Markov chain), 
the conditions of Th. \ref{thbo1}
are equivalent 
to the ones ensuring
Harris ergodicity. 
In Harris' framework, the conditions are on
the state 
space. In Borovkov's framework, the conditions are on the exogenous driving
sequence. This second approach is better suited for our problem.  On the one
hand, a direct analysis on the state
space appears to be almost inextricable. 
On the other hand, the 
renovating events will take a very convenient form because a
product 
of matrices is still a matrix (see Theorems \ref{th1}, \ref{th2}).

\paragraph{$\eta$-coupling}
Coupling and strong coupling, introduced 
above, are related to total variation convergence. 
We define now the notion of $\eta$-coupling. It is related to weak 
convergence.

\begin{defi}[$\eta$-coupling]
We consider a metric space $(E,d)$. We consider two sequences $\{X_n\}_{n \in
\N}$ and $\{Y_n\}_{n\in \N}$ defined on $E$. We say that there 
is $\eta$-coupling\footnote{The classical terminology is
$\eps$-coupling. We change it to $\eta$-coupling to
avoid confusions with the notation $\eps=-\infty$ of the $\R_{max}$ algebra.}
of  
these two sequences if for each $\eta >0$, one 
can find versions of $\{X_n\}$ and $\{Y_n\}$ defined on a common probability
space and an a.s. finite random time $N$ such that 
\[
n\geq N \Longrightarrow d(X_n,Y_n) \leq \eta\:.
\]
\label{eps} 
\end{defi}
The following proposition is shown in Asmussen \cite{asmu92}.

\begin{prop} 
We consider a sequence $\{X_n\}_{n \in \N}$ and a stationary sequence
$\{Y\circ \theta^n\}_{n\in \N}$ defined on the metric space $E$.
Let $\mu$ be the invariant distribution of $Y$. If there is $\eta$-coupling of the two sequences, then $\{X_n\}$ converges
weakly to $\mu$.
\label{asmu}
\end{prop}

\section{Presentation of the Results}
\label{se:prr}
Let us consider a probability space $(\Omega, {\cal F}, P, \theta)$. The
probability $P$ is stationary and ergodic with respect to the shift $\theta$.
We are interested in systems of the type: 
\[
\left\{ \begin{array}{ccc}
x(n+1)& = & A(n) \otimes x(n),\;\; n \in \N \\
x(0) & = & x_0 
\end{array} \right.
\]
where $ x(n)$ and $A(n)$ ($\forall n$) are finite, respectively $\R^k_{max}$
and 
$\R^{k \times k}_{max} $-valued, random variables. 
We are sometimes going to use the notation
$x(n,x_0)$ to 
emphasize the value of the initial condition. We will
consider models where the sequences $\{A(n), n\in \N\}$ are respectively {\bf
i.i.d}  or
{\bf stationary and ergodic} (i.e $A(n+1)=A(n)\circ \theta$).

\parag

We recall that we have defined first and second order
limits in Section \ref{se-intro}, Equations \eref{eq-1ord} and \eref{eq-2ord}.

We are going to recall results on first and second order limits for open
systems and first order limits for closed systems before
completing the picture by solving the problem of second order limits for
closed systems. 

\subsection{Results from Baccelli \protect\cite{bacc92} and Cohen
\protect\cite{cohe}} 
\label{sse:rfbacc}
For $x\in \R^k$ and $A
\in \R^{k\times k}_{max}$, we use
the notation $\|x\|_{\infty}=\bigoplus_{i=1}^{k} x_i$ and $\|A\|_{\infty}=\bigoplus_{i,j=1}^{k}
A_{ij}$.

\subsubsection{First order limits for closed systems}

\begin{theo}[Cohen \protect\cite{cohe}]
Let $\{A(n)\}$ be a stationary and ergodic sequence of 
matrices. We suppose that the matrix $A(0)$ has a fixed structure (see
Definition \ref{struc}), is
irreducible and verifies $P(A_{ij}(0)=\eps)=1$ or
$\eps < E(A_{ij}(0)) < +\infty$. 
There exists a constant $\lambda \in \R$ such
that, for all initial condition $x_0$ and for all $i\in \{1,\dots ,k\}$:
\[
\lim_n \frac{ x_i(n,x_0) }{n} =\lim_n E\left(\frac{ x_i(n,x_0) }{n}\right)
=\lambda,\;\;P-a.s. 
\]
The constant $\lambda$ is called the Lyapunov exponent of the stochastic
matrix $A(0)$. 
\label{first}
\end{theo} 

The basic idea is to use the inequality $\|A\otimes B\|_{\infty} \leq 
\|A\|_{\infty} \otimes \| B\|_{\infty}$ in order to apply 
Kingman subadditive ergodic theorem. \\

This definition of a  Lyapunov exponent is coherent with the one
of Theorem 
\ref{lyap}. Indeed, by Theorem \ref{thcyc}, for every irreducible and
deterministic matrix $A$, there 
exists $d$ and $M$ such that $\forall m \geq
M,\;A^{m+d}=\lambda^{d}\otimes A^m$, where $\lambda$ is the eigenvalue of $A$.
It implies that $\forall x_0\in \R^k_{max}, \lim_n A^nx_0/n=\lambda$.

\subsubsection{First order limits for open systems}

We suppose that matrices $A(n)$ have a fixed structure.
We decompose the graph of $A(0)$ 
into its maximal strongly connected
subgraphs (s.c.s.). If we replace each s.c.s. by one node, we obtain an associated
reduced graph which is acyclic. We associate with each node $\tilde{u}$ of the
reduced 
graph a constant $\lambda_{\tilde{u}}$ which is the Lyapunov exponent of the
corresponding s.c.s. in isolation, see Theorem \ref{first}. We denote
by $^{\bullet}\tilde{u}$ the
set of predecessors of $\tilde{u}$ (including $\tilde{u}$) in the reduced
graph. We have :

\begin{theo}[Baccelli \protect\cite{bacc92}]
Let $\{A(n)\}$ be a stationary and ergodic sequence of 
matrices. We suppose that $A(0)$ has a fixed structure. We suppose also that
$P(A_{ij}(0)=\eps)=1$ or $ \eps <E(A_{ij}(0))<+\infty,\forall i,j$.  Let us
consider $i \in 
\{1,\dots ,k\}$, $i$ belongs to the s.c.s. $\tilde{u}$.
\[
\lim_n \frac{ x_i(n,x_0) }{n} =\lim_n E\left(\frac{ x_i(n,x_0) }{n}\right) =
\bigoplus_{\tilde{v} \in \:^{\bullet}\tilde{u}} \lambda_{\tilde{v}},\;
P-a.s.\:.
\]
\label{first2}
\end{theo}

Intuitively, 
the dynamic of the system is imposed by the s.c.s.
having 
the smallest throughput (largest cycle time $\lambda$).

\subsubsection{Second order limits for open systems}
Matrices $A(n)$ 
have a fixed structure. In order to simplify the presentation of the
results, let us assume that the structure consists of two s.c.s. The general
case is completely similar. Up to a permutation of the coordinates, we have~:
\[
A(n)=\left(
\begin{array}{cc}
\tilde{U}(n) & \eps \\
\tilde{B}(n) & \tilde{A}(n)
\end{array}\right)\:.
\]
The block $\tilde{U}$ is a square matrix of size $I\times I$, irreducible. It
is interpreted as the input 
of our system. The block $\tilde{A}$ is a square matrix of size $(k-I)\times
(k-I)$ , irreducible. The block $\tilde{B}$ is the matrix of the
communications between the sources ($\tilde{U}$) and 
($\tilde{A}$). We suppose that the block 
$\tilde{U}$ in isolation has a unique stationary regime.
We have the following theorem.

\begin{theo}[Baccelli \protect\cite{bacc92}]
Let $u$ and $a$ be the Lyapunov exponents of $\tilde{U}$ and $\tilde{A}$
respectively (see Theorem \ref{first}). If $a<u$, 
there is a unique stationary regime 
for the SRS $\pi(x(n))$, regardless of the initial condition. Convergence to
the stationary regime occurs with strong coupling.
If $a>u$, then the differences of the form
\[
x_j(n,x_0)-x_i(n,x_0),\;i=1,\dots,I,\;\;j=I+1,\dots,k \:,
\]
tend to $+\infty$, $P-a.s.$, for all finite initial condition.
\label{bacc}
\end{theo}

If $u>a$, the sources which
are slower impose their pace. If $u<a$, 
everything happens asymptotically as if $\tilde{A}$ were in isolation.
 
\parag

{\bf Remark } In the previous theorem, we need the assumption that
$\tilde{U}$ 
in isolation has a 
unique stationary regime. Knowing if $\tilde{U}$ has a
unique stationary regime is precisely the problem which is going to be  
addressed in the following.
Then, to determine if there is a unique stationary regime for $\pi (x(n))$, we
have to use the results of Section \ref{sse:sgm} (applied to
$\tilde{U}$) together with  
the comparison of Lyapunov exponents (of
$\tilde{U}$ and $\tilde{A}$).

\parag

{\bf Remark } 
In the results above (Theorems \ref{first}, \ref{first2} and \ref{bacc}), the
assumption that a 
matrix 
$M(0)$ has a fixed structure and is irreducible can be weakened and replaced
by: 
\[
\lim_{n} P \left(\: M(n)\otimes M(n-1)\otimes \cdots \otimes M(0)
\: \mbox{irred. }\:\right)=1,\;\eps <E(M_{ij}(0)\:|\: M_{ij}(0) \neq \eps )
<+\infty\:. 
\]

\subsection{Preliminary results}
\label{sse:prr}

From now on, we concentrate on second order limits in the closed (i.e $A(n)$ is
$P$-a.s. irreducible) case. 
The 
limits
are expected to be random variables. We are interested in determining whether
the limiting distribution is unique.
Furthermore, we want to investigate the type 
of convergence to the limit.

\parag

We recall that $\pi$ is the canonical projection $\R^k
\stackrel{\pi}{\longrightarrow} \P\R^{k}$ (Def. \ref{de:pi}). It is clear
that the  
recursive 
equation $x(n+1)=A(n)x(n)$ defines a SRS (Def. \ref{srs}). It implies that
$\pi(x(n))$ is also a SRS. Indeed, let us
consider $x(n)$ and $x'(n)$ such that $\pi(x(n))=\pi (x'(n))$. We define
$x(n+1)=A(n)x(n)$ and 
$x'(n+1)=A(n)x'(n)$. It is straightforward that $\pi(x(n+1))=\pi (x'(n+1))$.
We write with some abuse of 
notation that $\{\pi(x(n)),\;n\in \N\}$ verifies
the recursive equation ``$\pi(x(n+1))=\pi A(n) \pi (x(n))$''\footnote{It
would be more rigorous to use different notations $\pi$ and $\tilde{\pi}$ for
the canonical projections in $\P\R^{k}$ and $\P\R^{k\times 
k}_{max}$ respectively. Then we would define more formally $\tilde{\pi}
(A(n))\pi 
(x(n))\equiv \pi(A(n)x(n))$.}.
%(identifying $A(n)$
%and its canonical projection in $\P\R^{k\times 
%k}_{max}$).
%We define isomorphisms $\phi_i ,\;i\in \{1,\dots,k\}$ between $\P\R_{max}$
%and $\R^{k-1}$ in the following way: 
%\[
%\phi_i=P_i \circ  \pi^{-1},\;\;{\rm where}
%\]
%\[
%P_i:\;\;\R^k \longrightarrow \R^{k-1}
%\]
%\[
%x \longmapsto 
%\begin{array}{c}
%x_1-x_i \\
%\vdots \\
%x_{i-1}-x_i \\
%x_{i+1} - x_i \\
%\vdots \\
%x_k -x_i
%\end{array}
%\]

\begin{lemm}[Reising \& al~\cite{Ral90}]
For $i \in \{1,\dots,k\}$, we define $z_i(n)=x_i(n)-x_{i}(n-1)$. 
We have $z_i(n)=F_i(\:A(n-1),\pi(x(n-1))\:)$, where $F_i$ is an absolutely 
continuous function.
\end{lemm} 

The sequence $\{A(n)\}$ being stationary by hypothesis,
it implies the following corollary. 

\begin{coro}
%A sufficient condition for $z_i(n)$, $i$ fixed, to couple in finite time
%(resp. to couple)  with a unique
%stationary 
%sequence is that $\pi(x(n))$ has the same property. (!!!!FAUX) A necessary and
A sufficient condition for $(z_1(n),\dots,z_k(n))'$ to
converge weakly (resp. in total variation) to a unique
invariant 
distribution, uniformly over initial conditions in $\P\R^{k}$, is that
$\pi(x(n))$ 
has the same property.
\label{co:co}
\end{coro}

This sufficient condition is not necessary as  demonstrated by the
following 
deterministic example.
\begin{exam}
\label{ex:det}
Let us consider 
\[
A=\left(
\begin{array}{cc}
e & -1 \\
-1 & e 
\end{array}
\right)\:.
\]
We have $A^+=A$, so $u_1=(e,-1)'$ and $u_2=(-1,e)'$ are eigenvectors of $A$. 
The set  
\[
\left\{u_{\lambda}=\lambda \otimes u_1
\: \oplus \: 
(1-\lambda)\otimes u_2,\;\lambda \in [0,1]\right\}\:,
\]
is the set of eigenvectors
of $A$, see Theorem \ref{vectprop}. There is a continuum of stationary regimes
for 
$\pi(x(n))$. For example, it is easy to check that for an initial condition
$u_{\lambda},\;\lambda \in [0,1]$, we have:  
\[
x_1(n,u_{\lambda})
-x_2(n,u_{\lambda}) =2\!\times \lambda -1\:.
\]
But on the other hand, we have a unique stationary regime for $z_i(n)$. As a direct consequence of the equality $A^2=A$, we have
$z_1(n)=z_2(n)=e,\;\forall n\geq 2$. \\
We can also easily build stochastic counter-examples of the same kind.
% based on the model of 
%Example \ref{ex:mult2}.
\end{exam}

{\bf Remark } The variables $\pi(x(n))$ depend only on the sequence
$\{\pi(A(n))\}$.
Therefore,
all the results on 
$\pi(x(n))$ 
would still be true under the weaker assumption that 
only the sequence $\{\pi(A(n))\}$ is stationary and ergodic. But, on the other
hand, the variables $z_i(n)$ depend on the sequence $\{A(n)\}$ and not only
on $\{\pi(A(n))\}$. Corollary \ref{co:co} would not be true under
the assumption that $\{\pi(A(n))\}$ is stationary and ergodic.
 
\parag

In the rest of the paper, 
we investigate the existence of a stationary regime for the SRS
$\pi(x(n))$, i.e. the existence of a finite r.v. 
$Z: \Omega \rightarrow \P\R^k$  such that\footnote{We will write
$Z\circ \theta = 
\pi A(0) Z$ with some abuse of notations.} 
\[
Z\circ \theta = \pi\left( A(0) \pi^{-1}(Z)\right)\:.
\]
We are interested by conditions
ensuring the uniqueness of the stationary regime and the convergence
of $\pi(x(n,x_0))$ toward it, for all $x_0\in \R^k$. In such cases, we say
that the model is {\it stable}. Two types of
convergence will appear, convergence with $\eta$-coupling and convergence with
coupling. They imply, respectively, weak convergence and total
variation convergence as recalled in \S \ref{se:bret}. 

\subsection{Stability of discrete models}
\label{sse:sdm}
Let $\left\{ A_l, \;l \in {\cal L} \right.$ or $ \left. l \in \N \right\}$, be
a finite or countable collection of irreducible 
matrices of size $k\times k$. We suppose that there 
exists a 
discrete probability law \{$p_{l}$\} such that $A(n, \omega)=A_{l}$ with
probability $p_l >0$.  

\begin{defi}[pattern, 1] $ $ \\
\label{de:patt}
A matrix $\tilde{A}$ is called a pattern of the random sequence $\{A(n),n\in
\N\}$ if: 
\begin{enumerate}
\item $\exists N \:|\: \tilde{A}=A_{u_{N-1}}\otimes\cdots\otimes A_{u_0}$ with
$u_0,\dots,u_{N-1} \in {\cal L}$ (or $\N$).
\item $P(A(N-1)\otimes \cdots \otimes A(0)=\tilde{A})>0$.
\end{enumerate}
If the sequence $\{A(n)\}$ is i.i.d. then the second condition is always
verified. 
\end{defi}

\begin{theo}
\label{th1}
The sequence of matrices $\{A(n)\}$ is i.i.d. 
If there exists a pattern of $\{A(n)\}$ whose critical graph has a unique
s.c.s. and whose cyclicity is 1 ({\bf scs1-cyc1} matrix), then 
$\{\pi(x(n))\}$ converges with strong coupling 
to a unique stationary regime. It implies total variation
convergence of $\{\pi(x(n))\}$ to its stationary distribution.
\end{theo}

\begin {proof}
Let $C=A_{u_{N-1}}\otimes\cdots\otimes A_{u_0}$ 
be a scs1-cyc1 pattern. We have,
using the cyclicity 1 
assumption  (Th. \ref{thcyc}), 
\[
\exists M \; | \; \forall m\geq M, \: C^{m+1}= \lambda C^{m}\:,
\]
where $\lambda$ is the Lyapunov exponent of $C$.
We conclude that for all initial condition $y$, $C^{M+1}y=C(C^My)=\lambda
\otimes C^M y$. It means that $C^My$ is an eigenvector of $C$. By the
assumption on the 
critical graph of $C$, there is a unique eigenvector (up to a constant)
denoted $y_0$ (Th. \ref{vectprop}).  We have
$C^{M}\otimes y=\mu(y)\otimes y_0,\;\;\mu(y)\in\R$, or equivalently
$\pi(C^{M}y)=\pi(y_0)$. We define  
\[
{\cal B}_i=\{ 
\omega | \: 
A(i+MN-1,\omega)\otimes\cdots\otimes A(i+1,\omega)\otimes A(i,\omega)=C^M \}\:.
\]
From the {\bf i.i.d.} assumption, it follows
that P(${\cal B}_i)\;>$ 0. On ${\cal B}_i$, and for all initial condition $y$,
we have:
\begin{eqnarray*}
x(i+MN)  & = & C^M\otimes x(i) \\
                     & = & \mu (x(i))\otimes y_0 \\
\Longrightarrow \;\; \pi (x(i+MN)) &=& \pi(y_0)\:.
\end{eqnarray*} 
%\begin{eqnarray*}
%x_u(i+M+1)-x_u(i+M) &=&  \lambda \:.
%\end{eqnarray*} 
We check that the sequence ${\cal B}_i$ is compatible
with the shift, i.e. ${\cal B}_i= {\cal B}_0\circ\theta^i$. We conclude that
${\cal B}_i$ is a stationary renovating 
event 
sequence for the SRS $\pi(x(n))$. We apply Borovkov's Theorem
(version 
\ref{thbo2} for the set $V=\P\R^k$, 
as we have obtained a sequence of renovating
events independent 
of the initial condition) and the uniqueness of
the stationary regime follows. 
\end {proof}

\begin{exam}[Cyclic Jackson Network 2] 
We consider a basic Cyclic Jackson Network with $k$ queues and $k$ customers.
Such a 
network can be represented by the (max,+) matrix given in \S \ref{sse:cjn},
Equation (\ref{eq-cjn}).
We suppose that the sequence of  service times
$\{\left(\sigma_1(n),\dots, \sigma_k(n)\right), \:n\in \N\}$ is i.i.d.  
However the random variables
$\sigma_1(n), \dots ,\sigma_k(n)$ need not be independent for a given $n$. 
We suppose also that the service times have a discrete support, i.e. can only
take a countable number of values. We are in the framework of Theorem
\ref{th1}. 
We conclude that a sufficient condition of stability is to find a scs1-cyc1
matrix among the (max,+) matrices corresponding to this network.
As a direct application of the result stated in Example
\ref{ex:jak1}, we obtain that a condition of stability is:
\[
P(\exists i \:|\: \sigma_i(n)>\sigma_j(n),\forall j\neq i) >0\; 
\mbox{or} \;
P(\sigma_1(n)=\sigma_2(n)=\cdots=\sigma_k(n))>0\:.
\]
\label{ex:jak2}
\end{exam}

\parag
We now give a version of Theorem \ref{th1} in the stationary and
ergodic case. 

\begin{theo}
\label{th2}
The sequence $\{A(n)\}$ is stationary and ergodic. We suppose that
there 
exists a finite pattern 
$C=A_{u_{N-1}}\otimes\cdots\otimes A_{u_0}$ which is 
scs1-cyc1 and of rank 1 (see Def. \ref{de:rank1}). 
We suppose that ${\cal B}=  \left\{\omega \: | \: A(N-1)A(N-2)\cdots 
A(1)A(0)=C \right\}$ is of strictly positive probability.
Then $\{\pi(x(n))\}$ converges with
strong coupling to a unique stationary regime.
\end{theo}

\begin {proof}
The proof resembles the one of Theorem
\ref{th1}. As $C$ is of rank 1, we
have (see 
Def. \ref{de:rank1}): $C^2=\lambda \otimes C$, where $\lambda$ is the Lyapunov
exponent 
of $C$. We conclude that:
\[
\forall y \in \R^k, \;C^2y=C(Cy)=\lambda Cy\:.
\]
It implies that $Cy$ is an eigenvector of $C$. As matrix $C$ is scs1, it has a
unique eigenvector $y_0$, up to a constant. On ${\cal B}_i={\cal B} \circ
\theta^i$, 
we have 
\begin{eqnarray*}
\pi(x(i+N))     &=& \pi \left( C x(i) \right)\\
                 & = & \pi(y_0)\:.
\end{eqnarray*}
We check that the sequence ${\cal B}_i$ is compatible with the shift and we
apply Borovkov's Theorem \ref{thbo2}. 
\end {proof} 

{\bf Remark} If the dependence between matrices is markovian, a
sufficient 
condition to get $P({\cal B}) >$ 0 is that 
$p(A_{u_i},A_{u_{i+1}})>0,\; \forall i=1,\dots,N-1$, where $p(.,.)$ is the
markovian 
transition kernel.

\parag
{\bf Remark} The conditions of this theorem are, of course, weaker than the
i.i.d. 
assumption of Theorem \ref{th1}. However we made an
additional assumption, namely that the pattern $C$ is of rank 1. This
assumption cannot be relaxed, as
shown by the
counter-example \ref{ex:erg1}.

\begin{exam}
\label{ex:erg1}
Let $\Omega=\{\omega_{1}, \omega_{2}\}$ be the probability space,
P=\{$\frac{1}{2}$, $\frac{1}{2}$\} the 
probability law, and $\theta$ the stationary and ergodic shift defined by:
$\theta 
(\omega_{1}) = \omega_{2}$ and $\theta
(\omega_{2}) = \omega_{1}$.
We consider
\begin{displaymath}
A=
\left(
\begin{array}{cc}
1-\eta & e \\
e & 1 
\end{array}
\right),\:
B=
\left(
\begin{array}{ccc}
1 & e \\
e & 1-\eta
\end{array}
\right),\;\; e< \eta <\!\!< 1\:.
\]
\[
\{A(n,\omega_{1})\}=A,B,A,B,\dots\;\;\;\{A(n,\omega_{2})\}=B,A,B,A,\dots\:.
\end{displaymath}
Both matrices $A$ and $B$ are scs1-cyc1 patterns of length 1. But
patterns which are scs1-cyc1 {\bf and} of rank 1 are for example $A^n$ or $B^n$ for
$n > [1/\eta]$. We have for any $n > [1/\eta]$, $P(\exists
N\:|\: A(N-1)\cdots A(0)=A^n)=P(\exists
N\:|\: A(N-1)\cdots A(0)=B^n)=0$. Hence the conditions of Theorem \ref{th2}
are not verified. In fact, there is a continuum of
possible periodic limits. Consider $x_0=(a,b)'$ with
$-1+\eta < a-b < 1-\eta$. Then the limit regime of $\pi(x(n))$ has a
state space which is either 
$\{\pi(a,b)',\pi(
a+\eta,b)'\}$ (with probability $\frac{1}{2}$), or $\{\pi(a,b)',\pi(
a,b+\eta)'\}$ (with probability $\frac{1}{2}$).
\end{exam}

\subsection{Stability of general models}
\label{sse:sgm}
In this section, we consider a general model where the coordinates of our
matrices have a support which can be discrete, 
absolutely continuous with respect to Lebesgue measure or a mixture of these
two cases.

\parag
We need the following definition, extending the notion of pattern we have
been using for finite models. 
Let $M$ be a 
deterministic matrix and $\eta>0$. We denote
by $B(M,\eta)$ the open ball of center $M$ and of radius $\eta$ for the supremum norm of $\R^{k \times k}$. We have $N\in B(M,\eta)$ iff
\index{$B(A,.)$ (ball)}
\[
\forall i,j,\: N_{ij} \in ] M_{ij}-\eta, M_{ij}+\eta [
\]

\begin{defi}[pattern, 2]
\label{de:pattern}\index{pattern}
Let $A$ be a random matrix. We say that $\tilde{A}$ is a pattern of $A$ if
$\tilde{A}$ is a deterministic 
matrix verifying
\[ 
\forall \eta >0,\: P\left\{ A \in B (\tilde{A},\eta) \right\}>0\:.
\] 
Equivalently, we can say that $\tilde{A}$ belongs to the support
of the random matrix $A$. It includes the cases where 
$\tilde{A}$ is an accumulation point (discrete case) or a boundary point
(continuous case) of the support.
\end{defi}

\begin{defi}[pattern, 3]
\label{de:pattern2}\index{pattern}\index{pattern!asymptotic}
Let $\{A(n),n\in \N\}$ be a sequence of random matrices. We say that the
deterministic matrix
$\tilde{A}$ is a pattern of the sequence $\{A(n)\}$ if 
\[
\exists N\: \mbox{s.t.}\:\forall \eta >0,\: P\left\{A(N-1)\otimes \cdots \otimes A(0) \in
B(\tilde{A},\eta)\right\}>0\:.
\]
Equivalently, we can say that
$\tilde{A}$ is a pattern (Def. \ref{de:pattern}) of the random matrix
$A(N-1)\otimes \cdots \otimes A(0)$. We say that $\tilde{A}$ is an
asymptotic pattern of $\{A(n)\}$ if 
\[
\forall \eta >0,\: \exists N_{\eta}\: \mbox{s.t.}\: P\left\{\pi (A(N_{\eta}-1)\otimes 
\cdots \otimes A(0) )\in
\pi (B(\tilde{A},\eta))\right\}>0\:.
\]

\end{defi}
{\bf Remark } This definition is coherent with the one given in Definition
\ref{de:patt} for a discrete model. Note that, for convenience reasons, 
asymptotic patterns are defined in the projective space $\P\R^{k\times k}$. 

\begin{theo}
\label{th3}
The matrices $A(n)$ are i.i.d. (resp. stationary and ergodic). 
We suppose that
there exists a 
matrix $C$ which is a pattern of 
$\{A(n)\}$ (see Def. \ref{de:pattern2}) and which is
scs1-cyc1 (resp. of rank 1).
Then the SRS $\{\pi(x(n))\}$ has a unique stationary regime $\{Z\circ
\theta^n\}$.  The convergence occurs with $\eta-$coupling. It 
implies weak convergence of $\pi(x(n))$ to its unique 
stationary distribution.
\end{theo}
\begin{proof}
We prove directly Theorem
\ref{conv3}, a stronger version of the result. 
It is done in Appendix, \S \ref{app:conv3}. 
\end{proof}

\begin{theo}
The sequence of matrices $\{A(n)\}$ is  i.i.d. or stationary and ergodic.
We assume that
there exists a set ${\cal C}$ of matrices such that~:
\begin{enumerate}
\item $\forall C \in {\cal C}$, $C$ is a matrix of rank 1.
\item $\forall C \in {\cal C}$, $C$ is a pattern of $\{A(n)\}$.
\item $\exists N \:|\:P\left( A(N-1)\cdots A(0)\in {\cal C} \right)>0$.
\end{enumerate}
Then $\{\pi(x(n))\}$ converges with strong coupling to a unique stationary
regime. 
\label{th4}
\end{theo}
The conditions of Theorem \ref{th4} are stronger than the ones of 
Theorem \ref{th3} as we require the patterns of rank 1 to be of
positive probability. On the other hand, we obtain a stronger type of 
convergence.
\begin{proof}
Let us define
${\cal B}=  \left\{\omega \: | \: A(N-1)A(N-2)\cdots 
A(1)A(0)\in {\cal C}  \right\}$ and ${\cal B}_i={\cal B} \circ
\theta^i$. Using that the matrices $C \in {\cal C}$ are of rank 1, we
obtain that, on the event ${\cal B}_i$, $\pi(x(i+N))$ is 
independent of the value of $\pi(x(i))$. It implies that 
$\{{\cal B}_n,n\in \N\}$ is a stationary sequence of renovating events. 
The result follows. 
\end{proof}

{\bf Remark } Theorems \ref{th1} to \ref{th4} do not require any aperiodicity
(Def. 
\ref{ape}) assumption on matrices $A(n)$. 
However, the pattern $C$ whose existence is essential in all of these theorems
is 
aperiodic. The condition ``scs1-cyc1'' implies 
aperiodicity.  

\begin{exam}[Cyclic Jackson Network 3]
\label{ex:jak3}
We consider the same i.i.d. model as in Example \ref{ex:jak2}. 
However, the
distributions of the service times are now general. 
We obtain, by using Theorems
\ref{th3} and \ref{th4}, the stability under the condition:\\
The support of the random vector $(\sigma_1(n), \dots,
\sigma_k(n) )$ contains at least one point such that:
\[
\exists i \:|\: \sigma_i(n)>\sigma_j(n),\forall j\neq i \:
\mbox{or such that}\; \sigma_1(n)=\sigma_2(n)=\cdots=\sigma_k(n)\:.
\]
If the previous condition occurs with strictly positive probability,
we obtain total variation convergence. Otherwise, we obtain weak 
convergence. 
Here  is a case with only weak convergence. We consider an i.i.d. CJN
with three queues and three customers. We assume that $\sigma_1=\sigma_2
=1$ and $\sigma_3$ is uniformly distributed over $[0,1]$. 
\end{exam}

\section{Converse Theorems}
\label{se:ct}
We are going to prove converses of Theorems \ref{th1}, \ref{th2}, \ref{th3} and \ref{th4}. We
will consider successively finite 
and general models of type:
``$x(n+1)=A(n)x(n)$'' where the matrices are of size $k\times k$ and are
$P-a.s$ irreducible. We will, moreover, always suppose that there 
exists a pattern whose projective diameter (Def. \ref{D}) is
finite,  i.e.
\begin{equation}
\label{eq-ass}
\exists n\; | \; P\left\{ {\tt D} (\:A(n)\cdots A(0)\:)<+\infty \right\}>0\:.
\end{equation}
It implies $\lim_n P\{ {\tt D} (\:A(n)\cdots A(0)\:)<+\infty \}=1$,
see the proof of Lemma \ref{le:loyn}.
This condition is very weak. In the i.i.d. case, it is enough
that there exists a pattern which is irreducible and
aperiodic. We comment further on this condition in Section \ref{se:wi}.

\subsection{Finite models in $\Q^{k\times k}_{max}$}
We consider a finite model: ``$x(n+1)=A(n)x(n)$'',
with $A(n)\in \left\{ A_l, \;l \in
{\cal L}=\{1,\dots,L\} \right\}$. We assume that the matrices 
are irreducible. We assume also that the matrices $A_l,l\in \cL,$
belong to $\Q^{k\times k}_{max}$, i.e. that their coordinates are
rational. 

\begin{theo}
The sequence of matrices $\{A(n)\}$ is  i.i.d. or stationary and ergodic.
When there is
a unique stationary regime, convergence to this regime
occurs with strong coupling. A
necessary and sufficient condition for the 
model to have a unique stationary regime is that there exists a matrix $C$
verifying 
\begin{enumerate}
\item $C$ is a matrix of rank 1
(Def. \ref{de:rank1}).
\item $C$ is a pattern of $\{A(n)\}$ (Def. \ref{de:pattern2}).
\end{enumerate}
\label{conv}
\end{theo}

\begin{proof} 
It is given
in Appendix, \S \ref{app:conv}. 
\end{proof} 

Theorem \ref{conv} is not true in general when the matrices $A_l,l\in \cL,$ 
belong to $\R^{k\times k}_{max}$, see the following 
counter-example.

\begin{exam}\label{ex-irra}
We consider the matrices 
\[
A=\left( \begin{array}{cc}
e & -1  \\
-1 & -\eta  
\end{array}\right),\;
B=\left( \begin{array}{cc}
-\eta' & -1  \\
-1 & e 
\end{array}\right)\:,
\]
where $0<\eta,\eta ' \ll 1$ and $\eta$, $\eta'$ are not co-rational,
i.e. $\eta /\eta' \not\in \Q$. 

\parag
Let  $u=(u_1,u_2)'\in \R^2$, we set $\psi(u)
=u_2-u_1$. 
We identify $\P\R^2$ and $\R$ using the function $\psi \circ \pi^{-1}$.
The matrices $A$ and $B$ are scs1-cyc1. Their respective and unique 
eigenvectors are $\psi(e_1)=-1$ and $\psi(e_2)=1$. For a 
vector $u=(u_1,u_2)'$ such that 
$\psi(u) \in [-1,1]$, we have
\begin{equation}\label{eq-effAB}
\psi( Au)=\max(\psi (u)-\eta,-1),\;\;
\psi (Bu)=\min(\psi(u)+\eta',1)\:. 
\end{equation}

We consider a Markov chain defined on the set 
$\psi^{-1}[-1,1] \subset \R^2$. The transition
probabilities are
\begin{itemize}
\item For $u$ such that $\psi(u) \in ]-1+\eta , 1 - \eta'[,$ 
$p(u,Au)=1/2,\;p(u,Bu)=1/2$.
\item For $u$ such that $\psi(u) \in [1 - \eta',1],$
$p(u,Au)=1$.
\item For $u$ such that $\psi(u) \in [-1,-1+\eta ],$ 
$p(u,Bu)=1$.
\end{itemize}
The behaviour of the Markov chain is illustrated in Figure \ref{fi-markov}.

\begin{figure}[hb]
\centerline{\input{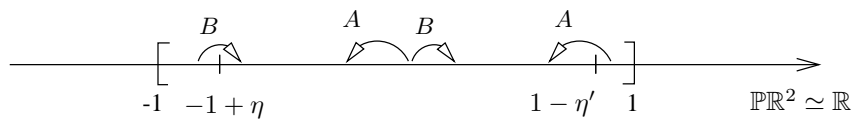}}
\caption{\sf Markov chain $\psi(X(n))$ on $\R$.}
\label{fi-markov}
\end{figure}

Let $X(n)$ be a realization of the Markov chain.
It is easy to check that this Markov chain is aperiodic.
Under the assumption $\eta /\eta' \not \in \Q$, one can prove using
classical arguments that the set $\{\psi(X(n)),n\in\N\}$ is $P$-a.s. 
dense in $[-1,1]$. It implies that the Markov chain is
$\nu$-irreducible where
$\nu$ is the Lebesgue measure on $\psi^{-1}[-1,1]$. 
Hence there exists a unique 
stationary distribution  $Q$ for the Markov chain.
It verifies $Q(\cA)>0$ for all event $\cA$ such that $\nu(\cA)>0$. 
For a complete treatment of Markov chains on continuous state spaces,
see Meyn \& Tweedie \cite{MeTw}.

Let us consider a 
stationary realization $X(n)$ of the Markov chain (i.e. 
$\forall n,\:P\{X(n)\in .\}=Q(.)$). 
We define
\begin{align}
A(n,\omega)=
\begin{cases}
A & \mbox{if $X(n+1,\omega)=AX(n,\omega)$,}\nonumber  \\
B & \mbox{if $X(n+1,\omega)=BX(n,\omega)$.}\nonumber
\end{cases} 
\end{align}
As $X(n)$ is stationary, it follows that $\{A(n)\}$ is a 
stationary and ergodic sequence. 

\parag
Let us consider the stationary-ergodic finite model ``$x(n+1)=A(n)x(n)$''
and $x(0)=x_0 \in \R^2$. Note that $\{x(n)\}$ is not 
a Markov chain anymore.

Let us consider a pattern $C=A_{n-1}\cdots A_0$ of $\{A(n)\}$, i.e.
$P\{A(n-1)\cdots A(0)= A_{n-1}\cdots A_0\}>0$. Let $x_0,\dots,x_n$ be
a corresponding path for the Markov chain $X(n)$, i.e.
\begin{eqnarray*}
x_0 \in \psi^{-1}]-1,1[,& & x_1= A_0 x_0,\dots , x_n= A_{n-1} x_{n-1},\: \\
& &\mbox{and}\: 
P\set{X(n)=x_n,\dots, X(1)=x_1}{X(0)=x_0}>0\:.
\end{eqnarray*}
Let us denote by $\uv{c}$ and $\ov{c}$ the 
minimal distances 
between $x_p,p\leq n$ and the extremal points of $\psi^{-1}[-1,1]$.
We have
$\uv{c}=\min_{p\leq n} (\psi(x_p)+1 )$ and 
$\ov{c}=\min_{p\leq n} (1-\psi(x_p))$.

It follows from \eref{eq-effAB} that 
\begin{equation}\label{eq-imag}
\psi (\Im (A_{n-1}\cdots A_0)) = [x_n -\uv{c},x_n + \ov{c}]\:,
\end{equation}
where 
$\Im(A)=\{Au,\: u\in \R^k\}$. 
From the definition of the Markov chain $X(n)$, it follows that
$\uv{c}>0, \ov{c}>0$. We conclude that $A_{n-1}\cdots A_0$ is not a 
rank 1 matrix. There exists no finite pattern of rank 1 for $\{A(n)\}$. 

\parag
On the other hand,  let us prove that there exists  asymptotic
patterns of rank 1 for $\{A(n)\}$. We define
$\uv{c}(n)=\min_{p\leq n} (\psi(X(p))+1 )$ and 
$\ov{c}(n)=\min_{p\leq n} (1-\psi(X(p)))$. 
As $\{\psi(X(n)),n\in\N\}$ is dense in $[-1,1]$, we obtain that
$\uv{c}(n)\rightarrow 0$ and $\ov{c}(n)\rightarrow 0$. 
Using
\eref{eq-imag}, we obtain that
${\tt D} (A(n)\cdots A(0)) \rightarrow 0,$ $P$-a.s.
We conclude following the lines of Theorem \ref{conv3}, \S \ref{app:conv3}. 
There is a unique 
stationary regime for the model. For an arbitrary 
initial condition,  we have $\eta$-coupling (weak convergence) with this
stationary regime.

\parag
To summarize, we have exhibited a finite model with a unique 
stationary regime and no coupling convergence. 
This type of behaviour is closely related to the 
non-finiteness of the projective semigroup $\pi<A,B>$, see Def. \ref{de-semi}.
\end{exam}

\subsection{General models}
We consider  a general model of type ``$x(n+1)=A(n)x(n)$''.
Stability no longer implies coupling in finite time.
It was illustrated by Example \ref{ex-irra}. Here is another example, for
an i.i.d. model.
\begin{exam}
\label{ex:erg2}
\[
A(n)=\left(
\begin{array}{cc}
U(n) & e  \\
e & U(n) 
\end{array}\right) \:,
\]
where $U(n)$ are i.i.d. random variables of uniform distribution over
$[0,1]$. There is a unique stationary 
regime for $\pi (x(n))$ which is $\pi (e,e)'$. We denote by $d(.,.)$ the
projective distance. For an initial condition $(y, e)'$ with 
$y \geq 1$, we have $d\left(x(n);(e,e)'\right)=\min_{p\leq n} U(p)$.
Thus convergence to $\pi (e,e)'$ occurs only asymptotically. 
There is no coupling but only 
$\eta$-coupling with the unique stationary regime.
\end{exam}

We can show the following results.

\begin{theo}
The sequence of matrices $\{A(n)\}$ is  i.i.d. or stationary and ergodic. The
necessary and sufficient condition for the 
model to converge with $\eta$-coupling to 
a unique stationary regime is the existence of
an asymptotic pattern $C$ of 
$\{A(n)\}$ of rank 1 (Def. \ref{de:pattern2}).
\label{conv3}
\end{theo}
\begin{proof} It is given in Appendix, \S \ref{app:conv3}.
\end{proof}

\begin{theo}
The sequence of matrices $\{A(n)\}$ is  i.i.d. or stationary and ergodic. The
necessary and sufficient conditions for the 
model to converge with coupling to 
a unique stationary regime are~:\\
There exists a set ${\cal C}$ of matrices such that~:
\begin{enumerate}
\item $\forall C \in {\cal C}$, $C$ is a matrix of rank 1.
\item $\forall C \in {\cal C}$, $C$ is a pattern of $\{A(n)\}$.
\item $\exists N \:|\:P\left( A(N-1)\cdots A(0)\in {\cal C} \right)>0$.
\end{enumerate}
We can say equivalently that we must have patterns of rank 1
but with 
strictly positive probability. 
\label{conv2}
\end{theo}

\begin{proof} We have already proved the sufficient part (Th. \ref{th4}).
We prove the necessary part of the theorem in Appendix, 
\S \ref{app:conv2}.
\end{proof} 

Convergence with $\eta-$coupling appears as a limiting case of
coupling in finite time. 
In a discrete model,
we will have only 
$\eta$-coupling when the set ${\cal C}$ of  scs1-cyc1 patterns is non empty
but is of 
probability 0. It means that the scs1-cyc1 patterns are only accumulation
points of the 
support. In a general model, we will 
have only  
$\eta$-coupling when the scs1-cyc1 patterns are isolated points
of the support 
(which implies that they 
are boundary 
points of the support). 

\begin{exam}
\label{ex:erg3}
To illustrate the previous remark, let us continue the analysis of Example
\ref{ex:erg2}. 
There is only one matrix
(in the projective space $\P\R^{k\times k}_{max}$)
verifying the first two conditions of Theorem 
\ref{conv2}. It is the matrix 
\[
\pi(C)=\pi \left(
\begin{array}{cc}
e & e  \\
e & e 
\end{array}\right) \:,
\]  
But condition 3. of Th. \ref{conv2} is not verified as $\forall N,
P\left(\pi(A(N-1)\cdots A(0))=\pi(C)\right)=0$. \\
Let us consider a slightly modified sequence of matrices $\{\tilde{A}(n)\}$
where the 
diagonal elements are two random variables $U(n)$ and $U'(n)$ defined 
on $[0,1]$ and such that
\[
P\left\{ U(n)>U'(n) \right\}>0 \;\;\mbox{or}\;\; P\left\{
U'(n)>U(n) \right\}>0\:. 
\]  
Now, we have scs1-cyc1 patterns with strictly positive probability and there is
coupling in 
finite time with the unique stationary regime. 
\end{exam}

\begin{exam}[Cyclic Jackson Network 4] 
\label{ex:jak4}
We consider the model of Example \ref{ex:jak3}. The condition
\[
P\left(\set{\exists i }{ \sigma_i(n)>\sigma_j(n),\forall j\neq i}\cup
\{\sigma_1(n)=\cdots=\sigma_k(n)\}\right) >0\:.
\]
is necessary
and sufficient for strong coupling convergence to a
unique stationary regime.
For i.i.d. Cyclic Jackson Networks,
the sufficient condition 
$P\left(\set{\exists i }{ \sigma_i(n)>\sigma_j(n),\forall j\neq i}\right)>0$ 
was obtained in
\cite{KaMa92}. The method of proof was completely
different, 
see the remarks at the end of Section \ref{sse:cjn}.
\end{exam}

\section{Without Irreducibility}
\label{se:wi}
We have supposed from the beginning that the matrices $\{A(n)\}$ were
irreducibles. 
The relaxation of the
irreducibility assumption is very important in terms of modeling power.
It enables us to 
consider, for example, the task graphs with random precedences introduced in
Section \ref{sse:tg}.

\parag

The irreducibility assumption
is used in 
Prop. \ref{decd}. But the only point we need to prove this
proposition is: 
``$\forall u \in \R^k,\; Au \in \R^k$'', i.e. if $u$ has only non-$\eps$
coordinates then $Au$ has the same property. So the only 
assumption we 
need on the matrices $\{A(n)\}$ is:\\

\hspace*{1cm} {\bf I} $\;\;\;\;\;\;\;\;\;\forall i,\:
P\left\{\exists j \;\mbox{s.t.} \; A_{ij}(0)>\eps \right\}=1\:.$ \\

The
irreducibility is also essential for the $\R_{max}$ spectral theory of Section
\ref{se:dst}.  
A reducible matrix $A\in \R^{k\times 
k}$ may
have several eigenvalues. Definition \ref{3def} and Theorem \ref{vectprop}
have to be reinterpreted by replacing the unique eigenvalue 
by the maximal 
eigenvalue. 
Theorem \ref{thcyc} is not true anymore.
But as far as the direct theorems
(\ref{th1}, \ref{th2}, \ref{th3} and \ref{th4}) are concerned, we use results of
the $\R_{max}$ spectral theory only for the pattern $C$ whose 
existence is critical for the proofs. These theorems are still valid if we state
that condition {\bf I} is verified and that there exists a 
pattern $C$ 
which is scs1-cyc1
{\bf and} irreducible.
 
\parag
Dropping the irreducibility assumption does not influence the converse results.
More precisely, the proofs of 
Theorems \ref{conv}, \ref{conv3} and \ref{conv2} are still valid. 
Only two conditions
need to
be verified: \\

\hspace*{1cm} {\bf I} $\;\;\;\;\;\forall i,\:
P\left\{\exists j \;\mbox{s.t.} \; A_{ij}(0)>\eps \right\}=1\:.$ \\
\hspace*{1cm} {\bf II} $\;\;\;\;$ $\exists n \;|\; P\left\{{\tt D}
(\:A(n)\cdots 
A(0)\:)<+\infty \right\} >0$. \\

Irreducibility $P-a.s.$ is not necessary to 
ensure that these conditions hold. We conclude that we can state our converse
results under the previous two minimal assumptions.

\parag

A counter-example shows that without condition {\bf II},
the uniqueness
of the stationary regime does not imply the existence of a
rank 1 pattern.

\begin{exam}
\label{ex:cond2}
Let $\Omega=\{\omega_{1}, \omega_{2}\}$ be the probability space, 
$P=\{\frac{1}{2}, \frac{1}{2}\}$ the
probability law, and $\theta$ the ergodic shift defined by: $\theta
(\omega_{1}) = \omega_{2}$ and $\theta
(\omega_{2}) = \omega_{1}$.
We consider
\begin{displaymath}
A=
\left(
\begin{array}{cc}
\eps & e \\
e & 1 
\end{array}
\right),\:
B=
\left(
\begin{array}{ccc}
e & e \\
1 & \eps
\end{array}
\right)\:.
\]
\[
\{A(n,\omega_{1})\}=A,B,A,B,\dots\;\;\;\{A(n,\omega_{2})\}=B,A,B,A,\dots\:.
\end{displaymath}
All patterns have an infinite projective diameter. Therefore,
condition {\bf II} is not verified. Nevertheless,
there is coupling in finite 
time with a unique periodic regime. More precisely, 
there is coupling of $\pi(x(n,u))$
to the periodic regime $\{\pi(e,e)',\pi(e,1)'\}$ 
and coupling occurs for $n>2\times [\bigoplus (u_{1}-u_2,u_2-u_1)]$. 
We conclude that there is coupling in finite time to a
unique stationary regime but no rank 1 pattern. 
Without condition {\bf II}, Theorem \ref{conv} is
not true anymore. 
\end{exam}

Another class of systems where condition {\bf II} is not verified is the class
of open systems studied by Baccelli in 
\cite{bacc92}. The results for this type of systems have been recalled in \S
\ref{sse:rfbacc}. In this case also, Theorem \ref{conv} fails to
be true. In such models, there are no  patterns which are scs1-cyc1 and
irreducible (matrices are non-irreducible with probability 1 !).
The good criterion to decide
the uniqueness of the stationary regime is the comparison between Lyapunov 
exponents, see Theorem \ref{bacc}. The computation of such exponents involves
the whole structure of 
the stochastic matrices $\{A(n)\}$, and not only an extracted 
pattern.

\parag
Condition {\bf II} is weak and will be verified in most cases.
For a discrete  i.i.d. model, for example, it is sufficient to have one pattern of finite length
$A_{u_N}\cdots A_{u_1}$ which is irreducible and aperiodic to verify it. For a general  i.i.d. model, it is sufficient to have
$P\left\{\:A(0) \right.$ irreducible and aperiodic$\left. \right\}>0$. In a stationary and ergodic framework, condition {\bf II} is a little bit
stronger, as shown by Example \ref{ex:cond2} where $P\left\{\:A(0) \right.$ irreducible and aperiodic$\left. \right\}=1$ and where
condition {\bf II} is not verified.

\parag

{\bf Remark }
For a general model which does not verify condition {\bf II}, we decompose the
model into its maximal sub-models verifying it. Then the complete analysis of
the system boils down to an analysis of the sub-models (using the results of
Section \ref{sse:sgm}) and of their interactions (using Theorem \ref{bacc} and
its generalizations, see \cite{bacc92}).

\section{Appendix}
\label{se:app}
\subsection{Loynes scheme}\index{Loynes scheme}\label{sse:ls}
Lemma \ref{le:loyn} is going to be used in several of the 
forthcoming proofs. Under an assumption of $\eta$-coupling of the
trajectories, we build a stationary regime using a Loynes' type 
construction. 

\begin{lemm}
\label{le:loyn}
We consider a general model ``$x(n+1)=A(n)x(n)$''
(see \S \ref{sse:sgm}). The sequence $\{A(n)\}$ is 
stationary and ergodic. We assume that there exists $N$ such that 
\[
P\left\{ {\tt D} (\:A(N)\cdots A(0)\:)<+\infty \right\} >0
\]
We assume also that $\forall x_0,y_0 \in 
\R^k,\:d(x(n,x_0),x(n,y_0))\rightarrow 0,\:P$ a.s. ($\eta$-coupling of the
trajectories). Then there exists a r.v. 
$Z:\Omega \rightarrow \P\R^k$ verifying
$Z\circ \theta=\pi(A(0))Z$.
The sequence $\{Z\circ \theta^n\}$ is the unique
stationary regime of the model.
\end{lemm}

\begin{proof}
We are going to show that the sequence
$\{\pi(A(-1)\cdots A(-n)u),\;n\in \N\}$, $u\in \R^k$,
has a simple limit in $\P\R^{k}$. 
The argument is an analog of the famous backward scheme proposed by 
Loynes in \cite{loyn} for G/G/1 queues. 
%We consider the following distance on $\P\R^{k \times k}_{max}$:
%\[
%{\tt d}(A,B)=\sup_{u \in \R^k} d(Au,Bu) \:.
%\]\index{${\tt d}(.,.)$}
%We recall that ${\tt D}(A)$ denotes the projective diameter of a matrix (see 
%Def. \ref{D}). We have: 
%\[
%{\tt d}\left( \: A(-1)\cdots A(-n), A(-1)\cdots A(-n)\cdots A(-n-p)\:\right)
%\leq {\tt D}(\:A(-1)\cdots A(-n)\:)\:. 
%\]%%%%%%%%%%%

We want to show that
${\tt D}(\:A(n)\cdots A(0)\:)\stackrel{n}{\rightarrow} 0,\;P-a.s$.
It is easy to see that the event 
\[
{\cal A}=\{\omega\:|\:\exists N,\;{\tt
D}(\:A(N)\cdots A(0)\:) \:<+\infty \}=\{\omega\:|\:\exists N,\forall n\geq N
,\;{\tt 
D}(\:A(n)\cdots A(0)\:) \:<+\infty \}
\]
is invariant by the translation shift.
Then by the ergodic Lemma, it is of probability 
0 or 1. We have made the assumption that $P({\cal A})>0$, hence 
$P({\cal A})=1$.

%In the i.i.d. case, a sufficient condition to get $P({\cal A})=1$ is that
%there exists a pattern which is irreducible and aperiodic. It 
%is under that very weak assumption that we are going to carry the proof (see
%Section \ref{se:wi} for more details).\\ 
%%% If $P({\cal A})=0$, then by Prop. \ref{easy} all finite products
%%% $A(n)\cdots A(0),\;n\in \N$ are reducible or 
%%aperiodic and then it is 
%possible to construct initial conditions $x_0,y_0$ such that
%$d(\:x(n,x_0),x(n,y_0)\:) \not\rightarrow 0$. 
% Let us precise the
%argument. Among the finite number of possible irreducible and non aperiodic
%structure of the communication graph, there is at 
%least one which will appear an infinite number of times, $P-a.s.$ in
%$\{A(n)\cdots A(0),\;n\in \N\} $. aperiodic But by Prop. 
%\ref{decd}, we have that  
%it was not the case, we would conclude using 
%Using the stationarity of the sequence $\{A(n),n\in \Z\}$, we have  
%\[
%\forall \eps>0,\;\forall u\in \R^k,\;\exists N
%So we have $P({\cal A})=1$. 
Using the stationarity of the sequence $\{A(n),n\in \Z\}$, we have that $
\exists N(\omega)$ such that
\[
{\tt D} \left(\: A(-1)\cdots A(-N)
\:\right)<+\infty\:.
\]
Then we can define the projective image of $A(-1)\cdots 
A(-N)$ which is a bounded subset of $\P\R^k$ and that we denote by $\Pi$.
The boundedness implies that
\[
c=\max_{v\in \Pi} d(e,v) < +\infty\:,
\]
where $e=(e,\dots ,e)'$. Let us define the vectors
\begin{equation}
\label{eq-bip}
c_1=(c,e,\dots,e)',c_2=(e,c,e,\dots,e)',\dots ,c_k=(e,\dots,e,c)'\:.
\end{equation}
It is
immediate that $\Pi$ is included in the convex hull of these vectors, i.e.
\[
\Pi \subset \left\{\pi(\alpha_1 \otimes c_1 \oplus \alpha_2 \otimes 
c_2 \cdots \oplus \alpha_k
\otimes c_k),\;\alpha_i \in \R\right\}\:.
\]
In the (max,+) algebra, we have the following property, 
for all $A\in \R_{max}^{k\times k},u,v \in \R_{max}^k, \; A(u\oplus
v)=Au\oplus Av$. It implies
\begin {equation}
\label{eq-semil}
\forall x \in \Pi, \; \pi (Ax) \in \left\{\pi(\alpha_1 \otimes Ac_1 \oplus 
\cdots \oplus \alpha_k
\otimes Ac_k),\;\alpha_i \in \R\right\}\:.
\end {equation}

%We will need the following
%lemma.  
%\begin{lemm}
%Let $K$ be a projectively bounded subset of $\R^k$. There exist vectors $u,v
%\in \R^k$ such that  
%\[
%\forall c\in K,\;d(u,c) \leq d(u,v)\;{\rm and}\;\;d(c,v) \leq d(u,v)\:.
%\]
%\end{lemm}
%\begin{proof} Let $\overline{K}$ be the closure of $K$. Let 
%$c_1,c_2 \in \overline{K}$ be such that  
%\begin {equation}
%\label{eq-closedset}
%d(c_1,c_2)=\sup_{c,c'\in K} d(c,c') \:.
%\end {equation} 
%We consider $u=c_1 + \alpha\times (c_1-c_2)$ and $v=c_2 + \alpha\times
%(c_2-c_1),\;\alpha>0$. 
%If we had $u\in K$, it would imply 
%$d(u,c_2)=d(c_1 + \alpha\times (c_1-c_2),c_2)=(1+\alpha)\times d(c_1,c_2)$
%which is impossible.  We prove in the same way that $v \not\in K$. 
%We
%have $d(u,v) =(1+2\alpha)\times d(c_1,c_2)$. We consider $c \in K$, we have: 
%\[
%d(u,c)\leq d(u,c_1)+d(c_1,c) \leq \alpha d(c_1,c_2) +
%d(c_1,c_2) < d(u,v) \:,
%\]
%\[
%d(v,c)\leq d(v,c_2)+d(c_2,c) \leq \alpha d(c_1,c_2) +
%d(c_1,c_2)< d(u,v) \:.
%\]
%\end{proof} 
%Let $x_0(\omega ),y_0(\omega )$ be two vectors verifying the previous lemma
%for the projective image of $A(-1)\cdots 
%A(-N)$. 

We fix $\eta>0$. Using the $\eta-$coupling assumption, we have that the random
variable $N'(\omega)$ is $P-a.s.$ 
finite, where $N'$ is defined by:
\[
N'=\inf\{n\;|\; d(x(n,c_i),x(n,c_j)) \leq \eta,\: \forall i,j \} \:.
\]

As both $N$ and $N'$ are $P-a.s.$ finite, we have
\[
\forall \delta >0,\;\exists L,L' \;:\;\; P\{N \leq L, N' \leq L'\}
\geq 
1-\delta \:.
\]
As a direct consequence of  \eref{eq-semil}, we have
on the event $\{N \leq L, N' \leq L'\}$~:
\[
{\tt D} \left( A(L'-1)\otimes \cdots \otimes A(0)A(-1)\otimes \cdots \otimes
A(-L) \right) \leq \eta \:. 
\]
We deduce, using the stationarity of $\{A(n)\}$, that
\begin{eqnarray*}
P\left\{ {\tt D} ( A(-1) \cdots A(-L-L') )\leq \eta \right\}&=&P\left\{ {\tt
D} ( A(L'-1) \cdots A(-L) )\leq \eta 
\right\} \\
 &\geq& 1-\delta \:.
\end{eqnarray*} 

It implies that the random variables ${\tt D} ( A(-1) \cdots A(-n) )$ converge
in probability to 0. But as ${\tt 
D}( A(-1) \cdots A(-n) )$ is pathwise decreasing, the convergence occurs also
$P-a.s.$ 

\parag

We have in particular, for all $u \in \R^k$, 
$d\left(\:A(-1)\cdots A(-n)u,A(-1)\cdots A(-n-p)u\right) \longrightarrow
0,\;P-a.s.$ It implies that $\{\pi(A(-1)\cdots 
A(-n)u)\}$ is a Cauchy sequence which converges. 
The limit does not depend on $u$. We denote it by $Z$. We have~:

%We denote its limit by
%$\pi(A^{\infty})$ where $A^{\infty}$ is a representative in $\R^{k\times
%k}_{max}$ 
%of  this limit. The representative $A^{\infty}$ is chosen to be normalized
%(Lyapunov exponent equal to $e$).
%The matrix $A^{\infty}$ depends on $\omega \in \Omega$ and
%has no 
%reason to be unique. %
%By 
%continuity of the projective distance, we have that $\forall u,v \in
%\R^k,\;d(A^{\infty}u,A^{\infty}v)=0$. Therefore $A^{\infty}$ is a 
%rank 1 matrix ($P-a.s.$). We define 
%$Z(\omega)=A^{\infty}u$ (independent of $u \in \R^k$). We have: 
\begin{eqnarray*}
Z \circ \theta & = & \lim_n \pi(A(0)A(-1)\cdots A(-n)u) \\
             & = & \pi A(0)\lim_n \pi (A(-1)\cdots A(-n)u) =\pi(A(0))Z\:.
\end{eqnarray*}
The sequence $\{Z \circ \theta^n\}$ is a stationary regime. Let us prove 
it is the unique one. 
We want to prove that
\begin{equation}
\forall x_0\in \R^k,\;\;d\left(x(n,x_0),Z\circ \theta^n\right) 
\stackrel{n\rightarrow +\infty}{\longrightarrow}  0\:,\;P-a.s. 
\label{eq-eq}
\end{equation}
As $Z$ is $P-a.s.$ finite, for all $\eta >0$, there exists a compact
$K\in \P\R^k$ such that $P\{ Z \in K \} >1- \eta$. We
proceed as above (Equation \eref{eq-bip}) in order to
define vectors $c_i,i=1,\dots,k$ such that 
\[
K \subset \left\{\pi(\alpha_1 \otimes c_1 \oplus \alpha_2 \otimes 
c_2 \cdots \oplus \alpha_k
\otimes c_k),\;\alpha_i \in \R\right\}\:.
\]
We have
\[
Z \in K \implies Z\circ \theta^p \in \left\{\pi(\alpha_1 
[A(p-1)\cdots A(0)c_1]\oplus 
\cdots \oplus \alpha_k
[A(p-1)\cdots A(0)c_k]),\;\alpha_i \in \R\right\}.
\]
Using the $\eta-$coupling of trajectories, we also have
\[
\forall c_i,\; d(x(n,x_0),x(n,c_i))\rightarrow 0\:.
\]
We conclude easily that
there is $\eta-$coupling of $\{\pi x(n,x_0)\}$ and $\{Z\circ
\theta^n\}$. 
We can apply Proposition
\ref{asmu}. There is weak convergence of $\{\pi x(n,x_0)\}$ to
the distribution of $Z$ and relation (\ref{eq-eq}) establishes the a.s. 
convergence of $\{\pi x(n,x_0)\}$ to $Z\circ \theta^n$. 
As a direct consequence, 
$\{Z\circ \theta^n\}$ is the unique stationary regime.
\end{proof}

\subsection{Proof of Theorem \protect\ref{conv} } 
\label{app:conv}
We are going to prove that the existence of a unique stationary regime
implies the existence of a pattern of rank 1 (Def. \ref{de:patt}). 
Using Theorem \ref{th2}, the proof will then be complete.

We need the following definition.
\begin{defi}
\label{de-semi}
Let us consider $A_1,\dots, A_p\in \R^{k\times k}_{max}$.
We denote by $<A_1,\dots A_p>$, the semigroup generated by these matrices 
and by $\pi<A_1,\dots A_p>$ its projection. We have
\begin{eqnarray*}
<A_1,\dots ,A_p>&=&\left\{ \left(A_{u_n}\cdots
A_{u_2}A_{u_1}\right),\;u_1,\dots,u_n \in \{1,\dots ,p\}, n \in \N
\right\}\:, \\
\pi\!\!<A_1,\dots ,A_p>&=&\left\{ \pi\! \left(A_{u_n}\cdots
A_{u_2}A_{u_1}\right),\;u_1,\dots,u_n \in \{1,\dots ,p\}, n \in \N
\right\}\:.
\end{eqnarray*}
\end{defi}

We consider the Euclidean space $(\P\R^{k\times k},\nopr{.})$
as introduced in Definition \ref{dist}. 
Next proposition was proved by Gaubert in \cite{gaub94b}.

\begin{prop}
\label{pr-gaub}
Let $A_1,\dots , A_p \in \Q^{k\times k}_{max}$. 
For all compact set $K$ of $(\P\R^{k\times k},\nopr{.})$, we have
$\pi<A_1,\dots,A_p>\cap K$ is finite. 
\end{prop}

Let us prove a lemma first.
\begin{lemm}
\label{le:boug}
We consider a finite model ``$x(n+1)=A(n)x(n)$''
with $A(n)\in \{A_1,\dots,A_p\}$ and $A_i\in \Q^{k\times k}_{max},i=1,\dots ,p$.
We suppose that there is a
unique stationary regime. It implies
\begin{equation}
\label{eq-eta}
d(A(n)A(n-1)\cdots A(0)x_0,A(n)A(n-1)\cdots A(0)y_0) \stackrel{n\rightarrow
+\infty}{\longrightarrow} 0,\;P-a.s\:.
\end{equation}
Equivalently, it implies $\eta$-coupling of the trajectories
corresponding to different initial conditions.
\end{lemm}

\begin{proof}
We assume that Equation \eref{eq-eta} is not verified. It implies,
using Proposition \ref{decd}, that there exists $x_0,y_0 \in \R^k$ and $c>0$
such that
\begin{equation}
\label{eq-boug}
P(\cA)>0,\;\; 
\cA=\{\lim_n d(A(n)A(n-1)\cdots A(0)x_0,A(n)A(n-1)\cdots A(0)y_0) >c\}>0\:.
\end{equation}
Let $S=<A_1,\dots , A_p>$ be the projective semigroup generated 
by the matrices of the model (Def. \ref{de-semi}).
For $x\in \R^k$, we define $S(x)= \{\pi(A x),\:A\in S\}$. 
We have that $\pi<A_1,\dots , A_p>\cap K$ is finite for all compact $K$
of $\P\R^{k\times k}$, Proposition \ref{pr-gaub}. 
It implies that $S(x)\cap K$ is finite for
all compact $K$ of $\P\R^k$. We conclude that $S(x)$ has no accumulation point
and verifies $\ov{S(x)}=S(x)$, where $\ov{S(x)}$ is the closure of $S(x)$
in $\P\R^k$. 

\parag
We want to apply Theorem \ref{anko}. It is required that the probability
space be a Polish space. 
In order to fulfill this,
we consider 
the canonical probability space 
consisting of one-sided infinite sequences of matrices $\{A_1,\dots A_p\}$, 
i.e.
\[
\Omega=\{(A_{u_0}, A_{u_1},\dots, A_{u_n},\dots),\; u_i\in \{1,\dots, p\} \}\:.
\]
We recall 
that we made the assumption \eref{eq-ass}, which implies
\[
\lim_n \:P\left\{ {\tt D} (\:A(n)\cdots A(0)\:)<+\infty \right\}=1\:.
\]
It implies that for all $\eta>0$, there exists $N\in \N$ 
and $K$, a compact set of $\P\R^{k\times k}$ such that
\[
\forall n\geq N,\:P\left\{ \pi(A(N)\cdots A(0)) \in K\right\} > 1-\eta\:.
\]
There exists a compact $K'$ (which depends on $x$)
of $\P\R^k$ such that 
\[
\{\pi(A(N)\cdots A(0)) \in K\}\implies 
\{\pi(A(N)\cdots A(0)x) \in K'\}\:.
\]
We conclude that the sequence
$\{\pi(x(n,x))\}$ is tight in $\P\R^k$. 
It implies that it is tight in $\ov{S(x)}=S(x)$. 
We can view $\pi(x(n,x))$ as
a SRS defined on $S(x)$ only. 
Applying Theorem \ref{anko}, we obtain that,
for all $x$, there exists a
stationary distribution $Q_{x}$ defined on $\Omega \times S(x)$.

\parag
Let us consider the initial conditions $x_0$ and $y_0$ as defined in
\eref{eq-boug}. It is a-priori possible to have $S(x_0)\cap S(y_0)
\neq \emptyset$. As a consequence, one cannot 
rule out that $Q_{x_0}=Q_{y_0}$. 
We are going to prove that there exists
$\alpha \in \R$ such that $S(x_0) \cap S(\alpha x_0 \oplus y_0)= \emptyset$. 
It will provide
two stationary
distributions $Q_{x_0}\neq Q_{\alpha x_0 \oplus y_0}$, which contradicts the 
uniqueness of the stationary regime. 

\parag

We work on the event $\cA$, see \eref{eq-boug}. We have $d(x(n,x_0),x(n,y_0))>c$ for all $n$. 
Let $x,y \in \R^k$ be two different points. Then there exists an
open interval $]\underline{\lambda},\overline{\lambda}[$ such that
\[
\overline{\lambda}-\underline{\lambda}  = d(x,y),\; \{\overline{\lambda}x \oplus y,
\underline{\lambda}x \oplus y\}=\{x,y\}\:,
\]
\[
\forall \lambda\neq\lambda' \in [\underline{\lambda},\overline{\lambda}],\:
\lambda x \oplus y \neq \lambda' x \oplus y\:.
\]
The proof is straightforward.
We consider
the (random) intervals 
$]\underline{\lambda}(n),\overline{\lambda}(n)[$ defined as above for
the couples
of points $\{x(n,x_0),x(n,y_0)\}$.
For any $A\in \R^{k\times k}_{max}, x,y \in \R^k_{max}$
and $\lambda \in \R$, we have $A(\lambda x \oplus y)= \lambda Ax \oplus Ay$. As a
consequence, the sequence $]\underline{\lambda}(n),\overline{\lambda}(n)[$ 
is decreasing. Let $\underline{\lambda}$ and $\overline{\lambda}$ be 
the limits of $\underline{\lambda}(n)$ and $\overline{\lambda}(n)$.
On
the event $\cA$, we have $\overline{\lambda}- \underline{\lambda}>c$ 
(see \eref{eq-boug}).

\parag
We define the sets
\[
\Lambda(n)=
\set{\lambda}{ \pi(\lambda x(n,x_0) \oplus
x(n,y_0)) \in S(x_0)},\; \Lambda =\bigcup_{n\in \N} \Lambda(n)\:.
\]

Let $x,y,z\in \P\R^k$ be three different points. It is immediate to
prove that there exists a unique $\lambda \in \R$ such that 
$z=\lambda x\oplus  y$.
As a consequence, the sets $\Lambda(n)$ are countable and $\Lambda$ is countable. 
It implies that 
the set
$]\underline{\lambda},\overline{\lambda}[\setminus \Lambda$ is non-empty on $\cA$.
For all $\lambda \in ]\underline{\lambda},\overline{\lambda}[\setminus 
\Lambda$, 
we have,
by definition of $\Lambda$, that $S(\lambda x_0 \oplus y_0)\cap S(x_0)=
\emptyset$. The conclusion follows.
\end{proof}

{\bf Remark } The proof does not work 
when matrices $A_1,\dots ,A_p$ belong to $\R^{k\times k}_{max}$. 
In this 
case, it is possible to have $\ov{S(x)}\neq S(x)$.  
In the model detailed in Example \ref{ex-irra},
all the sets $S(x)$ are dense in the interval $[-1,1]$
(as a classical consequence of the assumption $\eta/\eta' \not\in \Q$). 
It implies that $\overline{S(x)}= [-1,1],\forall x$.
The stationary distributions $Q_x$ are all defined on the same set, $\Omega \times [-1,1]$,
which prevents the previous proof from working. 

\parag

We want to prove the existence of a rank 1 pattern of $\{A(n)\}$ 
(Def. \ref{de:patt}).
%We want to prove that the convergence 
%obtained in Lemma \ref{le:boug} occurs in finite time. We suppose it is
%not the case, i.e. 
%there exist $x_0,y_0 \in \R^k$ such that $P(d(x(n,x_0),x(n,y_0)) \neq
%0,\;\forall n)>0$. By using Theorem \ref{th2}, 
%it implies that there exists no rank 1 pattern of $\{A(n)\}$ 
%(Def. \ref{de:patt}).%
There exists a r.v. $N$ such that 
$A(N)\cdots A(0)_{ij}>\eps,\forall i,j$ (consequence
of Equation \eref{eq-ass}). 

It follows from the ergodic Lemma, that the set
\begin{equation}
\label{eq-cI}
\cI=\set{n}{n\geq N,\:A(n)\cdots A(n-N)=A(N)\cdots A(0)}
\end{equation}
is infinite, $P$-a.s. Let $\sigma :\N\rightarrow \N$ be the strictly increasing
function such that $\cI=\{\sigma(0),\sigma(1),\dots\}$. We define the 
subsequence $\{B(n)=A(\sigma(n))A(\sigma(n)-1)\cdots A(0),~n \in \N\}$. 
%We are going
%to prove that the matrices $\pi(B(n))$ belong to a compact 
%of $\P\R^{k\times k}_{max}$. 
The matrices $B(n)$ can be written under the 
form $B(n)=A(N)\cdots A(0)\tilde{B}(n) A(N)\cdots A(0)$ for $n \geq 3$. We have
\begin{eqnarray}
\max_{ij} B(n)_{ij}& \leq &\max_{ij} A(N)\cdots A(0)_{ij}  \otimes \max_{ij} 
\tilde{B}(n)_{ij}
\otimes \max_{ij} A(N)\cdots A(0)_{ij} \nonumber\\
  & \leq & \max_{ij} A(N)\cdots A(0)_{ij} \otimes \tilde{B}(n)_{uv} \otimes
\max_{ij} A(N)\cdots A(0)_{ij}\:, \label{eq-cest1}
\end{eqnarray}
for some indices  $u,v$ belonging to the argmax in $\max_{ij} \tilde{B}(n)_{ij}$.
We also have
\begin{eqnarray}
\forall i,j,\;B(n)_{ij} &\geq &A(N)\cdots A(0)_{iu} \otimes \tilde{B}(n)_{uv} 
\otimes A(N)\cdots A(0)_{vj} \nonumber\\
\min_{ij} B(n)_{ij} &\geq & \min_{ij} A(N)\cdots A(0)_{ij} 
\otimes \tilde{B}(n)_{uv} 
\otimes \min_{ij} A(N)\cdots A(0)_{ij}\:.\label{eq-cest2}
\end{eqnarray}
We consider the Euclidean space $(\P\R^{k\times k}, \nopr{.})$ where $\nopr{.}$
is the norm introduced in Definition \ref{dist}. It follows from \eref{eq-cest1} and \eref{eq-cest2} that
\begin{eqnarray*}
\nopr{B(n)}= \max_{ij} B(n)_{ij} -\min_{ij}B(n)_{ij} &\leq& 2 \times (\max_{ij} A(N)\cdots A(0)_{ij} - \min_{ij} A(N)\cdots A(0)_{ij} )\\
  &=& 2 \times\nopr{A(N)\cdots A(0)}\:.
\end{eqnarray*}
It implies that the sequence $\{\pi(B(n))\}$ belongs to a compact 
of $(\P\R^{k\times k}, \nopr{.})$. Hence there exists a strictly increasing
function $\sigma :\N\rightarrow \N$ such that $\pi(B(\sigma(n)))$
is converging. Let $A_{\infty}$ be a representative (in $\R^{k\times k}$)
of the limit. 
%The matrix $A_{\infty}$ is chosen to be normalized 
%(Lyapunov exponent equal to $e$).
By 
continuity of the projective distance, we have that $\forall u,v \in
\R^k,\;d(A_{\infty}u,A_{\infty}v)=0$. Therefore $A_{\infty}$ is a 
rank 1 matrix. 

\parag
As the products $\{\pi(A(n)\cdots A(0))\}$ 
can only take a finite number of values in compact sets
(Proposition \ref{pr-gaub}),
it implies  that 
%The set of rank 1 matrices can be defined as 
%$\set{A}{\pi(A^2)=\pi(A)}$. As the function $\pi(.^2)-\pi(.)$ is
%continuous, this set is closed 
%in $\R^{k\times k}_{max}$ (or equivalently in
%$\P\R^{k\times k}_{max}$). We conclude that 
the limit matrix $A_{\infty}$
is attained in finite time. More precisely, there exists $N$ such that
\[
\forall n\geq N, \:\pi\left(B(\sigma(n))\right)=\pi (A_{\infty})\:.
\]
The matrix $B(\sigma(N))$ is a rank 1 pattern
for $\{A(n)\}$. It concludes the proof.
\cqfd

\subsection{Proof of Theorem \protect\ref{conv3} } 
\label{app:conv3}
We first prove the necessary part of the Theorem, i.e. $\eta-$coupling
with a unique stationary regime implies the existence of an 
asymptotic pattern.

Let $Z\circ \theta^n$ be the unique stationary regime. We have for all
$x_0,y_0 \in \R^k$,
\[
d(x(n,x_0),Z\circ \theta^n)\rightarrow 0,\: d(x(n,y_0),Z\circ \theta^n)\rightarrow 0 ~\implies ~ d(x(n,x_0),x(n,y_0))\rightarrow 0\:.
\]
%Using the triangular inequality, we deduce that 
%$d(x(n,x_0),x(n,y_0))\rightarrow 0$. 

We have assumed that 
$\exists N$ such that $P\{{\tt D} (A(N-1)\cdots A(0))<+\infty\}>0$, see 
Equation \eref{eq-ass}, Section \S \ref{se:ct}. 
Let $K \in \R$ be such that $P\{{\tt D} (A(-1)\cdots A(-N))<K\}>0$. 
It implies that there exists $K'$ such that
$P\{\nopr{A(-1)\cdots A(-N)}<K'\}>0$. 
Let us denote 
\[
\cE_0=\set{\omega}{\:\nopr{A(-1)\cdots A(-N)}<K'}\:.
\]
It follows from the stationary-ergodic assumption, that there exists 
a minimal $n_1>1$ such that
\[
P\{\cE_1\}>0,\; \cE_1=\cE_0\cap \{\nopr{A(-n_1)\cdots A(-n_1-N+1)}<K'\}\:.
\]
We define in the same way an increasing sequence $n_p>\cdots >n_2>n_1$ and
a decreasing sequence of events $\cE_p \subset \cdots \subset \cE_2 \subset
\cE_1$ verifying 
\[
P\{\cE_p\}>0,\; \cE_{p}=\cE_{p-1}\cap
\{\nopr{A(-n_p)\cdots A(-n_p-N+1)}<K'\}\:.
\]
On the event $\cE_p, p\geq 1$, we have 
\[
\nopr{A(-1)\cdots A(-n_p-N+1)} < 2\times K'\:.
\]
The proof is exactly similar 
to the one proposed in the proof of Theorem 
\ref{conv} (\S \ref{app:conv}, Equation \eref{eq-cI} and after). 
\index{$\cB(A,.)$ (projective ball)}
Let $\cB(E,K')$ denote the open ball of $(\P\R^{k\times k},\nopr{.})$ of
center $\pi(E), E_{ij}=e,\forall i,j$ and of radius $K'$. 
For all $p$, we
choose a deterministic matrix $B_p$ belonging to $\cB(E,K')$
and verifying 
\begin{equation}
\label{eq-verge}
P\left\{\{\cE_p\}\cap\{\pi A(-1)\cdots A(-n_p-N+1) \in \cB(B_p, 
\frac{1}{p})\}\right\}>0\:.
\end{equation}
As the matrices $\{B_p,p\in \N\}$ belong to a compact, there exists a
subsequence $\{B_{\sigma(p)}\}$ which converges to a limit $B_{\infty}$. 
We have (see the proof of Lemma \ref{le:loyn}) that
${\tt D }(A(-1)\cdots A(-n)) \rightarrow 0,\:P-a.s.$. We conclude that
$B_{\infty}$ is a rank 1 matrix. 

We fix $\eta >0$. Let $C$ be such that $\forall p>C$, we have
$\nopr{B_{\infty}-B_{\sigma(p)}}\leq \eta/2$. For $p>\max (C,2/\eta)$, we have
$\cB(B_{\sigma(p)},1/p) \subset \cB(B_{\infty},\eta)$. It implies
\begin{eqnarray*}
&&P\{\pi A(-1)\cdots A(-n_{\sigma(p)}-N+1)  \in \cB(B_{\infty},\eta)\}\geq   \\
&& \;\;\;\;\;\;\;  P \left\{\{\cE_{\sigma(p)}\}\cap\{\pi A(-1)\cdots A(-n_{\sigma(p)}-N+1) \in \cB(B_{\sigma(p)}, 
\frac{1}{p})\}\right\}>0\:.
\end{eqnarray*}
It means precisely that $B_{\infty}$ is an asymptotic
pattern of $\{A(n)\}$, see Definition \ref{de:pattern2}.

%We can apply Lemma
%\ref{le:loyn}. We consider a given $\omega
%\in \Omega$. It follows from the proof
%of Lemma \ref{le:loyn}, see \S \ref{sse:ls}, that 
%${\tt d} (A(-1)\cdots A(-n),A(-1)\cdots A(-n-p))\stackrel{n}{\rightarrow} 0$. 
%The difficulty is that ${\tt d}(.)$ is not a
%distance on $\P\R^{k\times k}$. Let us view $\P\R^{k\times k}$ as 
%the vector space $\P\R^{k^2}$ equipped with
%the coordinatewise operations $\oplus$ and $\otimes$. 
%The neutral element of $\P\R^{k^2}$
%is the matrix $\pi(E),\: E_{ij}=e,\forall i,j$. We consider the projective norm 
%$\nopr{.}$ on $\P\R^{k^2}$ as introduced in Definition \ref{dist}.
%We have
%\[
%\nopr{A}=\max_{ij} A_{ij} -\min_{ij} A_{ij}\:.
%\]%%%%%%%%%

\parag

Let us prove the sufficient part of the theorem. We assume that there exists
a deterministic 
matrix $\tilde{A}$ which is a rank 1 asymptotic pattern of $\{A(n)\}$.
We want to prove the $\eta$-coupling convergence of $\pi(x(n))$ to a 
unique stationary regime. 

We fix $\eta >0$. Let $N_{\eta}$ be such that 
\begin{equation}
\label{eq-triv}
P\left\{\pi A(N_{\eta}-1)
\cdots A(0) \in
\cB(\tilde{A},\eta)\right\}>0\:.
\end{equation}
Using the ergodic Lemma, we have
\[
P\set{\exists i \geq 0}{\pi A(N_{\eta}-1+i)  
\cdots   A(i) \in
\cB(\tilde{A},\eta)}=1\:.
\]
Let $u$ be the unique eigenvector of the rank 1 matrix $\tilde{A}$
and 
$\cB(u,\eta)$ the ball of center $\pi(u)$ and radius $\eta$ in $\P\R^k$. 
We have
that for all $x_0 \in \R^k$,
\[
\{ \pi(x(n,x_0)) \in \cB(u, \eta) \} \subset 
\{\pi A(n-1)  
\cdots   A(n-N_{\eta}) \in
\cB(\tilde{A},\eta)\}\:.
\]
In particular, it implies that $\forall x_0,y_0 \in \R^k$ and $n$ large enough,
\[
\{d(x(n,x_0),x(n,y_0)) <\eta\} \subset \bigcup_{N_{\eta} \leq p \leq n}
\{\pi A(p-1)  
\cdots   A(p-N_{\eta}) \in
\cB(\tilde{A},\eta)\}\:.
\]
We deduce that
$P\{d(x(n,x_0),x(n,y_0)) <\eta\}\rightarrow 1$. We conclude by using 
Lemma \ref{le:loyn} (the existence of $n$ such that $P\{ {\tt D} (A(n)\cdots
A(0)<+\infty\}>0$ comes from Equation \eref{eq-triv}).

\cqfd

\subsection{Proof of Theorem \protect\ref{conv2}}
\label{app:conv2}
We want to prove that the conditions given in Theorem \ref{conv2} are
necessary. We suppose that our model couples in finite time with a unique
stationary regime, uniformly over initial
conditions in $\R^k$. 
%\marginpar{a montrer!}
%We have to assume that there is at least a pattern with an irreducible and
%aperiodic pattern (in the same way than in the proof of Theorem
%\ref{conv}).
Let us prove a lemma first.

\begin{lemm}
If there is a unique stationary regime 
for $\pi (x(n))$, coupling in finite time uniformly over
initial conditions in $\R^k$ implies strong coupling in finite time uniformly
over initial conditions in $\R^k$.
\end{lemm}

\begin{proof} Let $\{ Z \circ \theta^n \}$ be the unique stationary regime with
which the SRS $\pi(x(n))$ couples.
We consider the event:
\[
{\cal Y}_n= \left\{ \omega \; | \;\pi( x(n,x_0) ) \circ \theta^{-n}
\omega=Z \omega,\;\; \forall x_0 \in \R^k \right\}\:.
\]
The assumption of coupling in
finite time, uniformly over $\R^k$, may be written~:
\[
P({\cal Y}_n) \stackrel{n\rightarrow +\infty}{\longrightarrow} 1 \:.
\]
Here we implicitly use the assumption that the projective image of
$A(-1)\cdots A(-n) $ is asymptotically bounded (see 
Equation \eref{eq-ass}).
Let us consider $\omega \in {\cal Y}_n$ and $p$ an integer
$>0$, we have:
\begin{eqnarray}
\pi (x(n+p,x_0)) \circ  \theta^{-n-p} \omega & = &\pi \left(\: x(n,
x(p,x_0)\circ 
\theta^{-p})\; \right) \circ \theta^{-n} \omega \label{eq-st1}\\
 & = & Z \omega \;\;\;\;({\rm as } \; \omega \in {\cal Y}_n) \:.\label{eq-st2}
\end{eqnarray}
The passage from \eref{eq-st1} to \eref{eq-st2} uses
the fact that coupling occurs uniformly over
initial conditions. We have:
\[
{\cal Y}_n=\left\{ \omega \; | \;\pi( x(n+p,x_0) ) \circ \theta^{-(n+p)}
\omega=Z \omega,\;\;\forall p \geq 0 ,\;\; \forall x_0 \in \R^k \right\}\:,
\]
and
\[
P({\cal Y}_n) \stackrel{n\rightarrow +\infty}{\longrightarrow} 1 \:.
\]
This is exactly the definition of strong coupling (Def. \ref{scoup}).
\end{proof} 

We can now use the converse Theorem \ref{thbo2c}. There exists a stationary
sequence of 
events $\left\{{\cal A}\circ
\theta^n \right\}$ which is renovating for the SRS $\{\pi\left( \: x(n,x_0) \:
\right)\},\;\forall x_0 \in \R^k$, and verifies $P({\cal A})>0$. Let $m$ be
the common length and $\Phi$ the common function of these renovating events.
We have, on ${\cal A}$:
\[
\pi(x(m))=\Phi \left( A(m-1),\dots,A(0) \right),\;\; \forall x_0 \in \R^k\:.
\]
But we also have:
\[
x(m)=A(m-1)\otimes\cdots\otimes A(0)\otimes x_0,\;\; \forall x_0 \in \R^k \:.
\]
We conclude that, on ${\cal A}$, $\pi \left(A(m-1,\omega)\otimes\cdots\otimes
A(0,\omega)\otimes x_0\right)$ is independent of $x_0$. It implies 
that $C=A(m-1,\omega)\otimes\cdots\otimes A(0,\omega) $ is a 
matrix of rank 1. \cqfd

\parag
\parag

{\bf Acknowledgment } I would like to thank Fran\c{c}ois Baccelli who
introduced me to this problem. 
F. Baccelli gave me also many ideas and suggestions which appear in this
paper. I am also grateful to Serguei Foss, St\'ephane Gaubert 
and Philippe Bougerol for 
several fruitful talks on the topic. At last, the careful comments of an
anonymous referee have greatly helped improving the presentation of this paper.

%% \small\bibliography{mairesse}

\begin{thebibliography}{}

\end{thebibliography}


\begin{thebibliography}{10}

\bibitem{AnKo}
V.~Anantharam and T.~Konstantopoulos.
\newblock Stationary solutions of stochastic recursions describing discrete
  event systems.
\newblock In {\em Proc. 33rd Conf. on Decision and Control}, volume~2, pages
  1481--1486, Lake Buena Vista, FL, 1994.

\bibitem{asmu92}
S.~Asmussen.
\newblock On coupling and weak convergence to stationarity.
\newblock {\em Annals of Applied Probability}, 2(3):739--751, 1992.

\bibitem{bacc92}
F.~Baccelli.
\newblock Ergodic theory of stochastic {P}etri networks.
\newblock {\em Annals of Probability}, 20(1):375--396, 1992.

\bibitem{BCOQ}
F.~Baccelli, G.~Cohen, G.J. Olsder, and J.P. Quadrat.
\newblock {\em Synchronization and Linearity}.
\newblock John Wiley \& Sons, New York, 1992.

\bibitem{BaLi92a}
F.~Baccelli and Z.~Liu.
\newblock On a class of stochastic recursive equations arising in queueing
  theory.
\newblock {\em Annals of Probability}, 21(1):350--374, 1992.

\bibitem{bamb}
N.~Bambos.
\newblock On closed ring queueing networks.
\newblock {\em J. Appl. Prob.}, 29:979--995, 1992.

\bibitem{boro84}
A.~Borovkov.
\newblock {\em Asymptotic Methods in Queueing Theory}.
\newblock John Wiley \& Sons, New York, 1984.

\bibitem{boro86}
A.~Borovkov.
\newblock Limit theorems for queueing networks. {I}.
\newblock {\em Theory Prob. Appl.}, 31:413--427, 1986.

\bibitem{boro88}
A.~Borovkov.
\newblock Limit theorems for queueing networks. {II}.
\newblock {\em Theory Prob. Appl.}, 32:257--272, 1988.

\bibitem{BoFo92}
A.~Borovkov and S.~Foss.
\newblock Stochastically recursive sequences and their generalizations.
\newblock {\em Siberian Adv. in Math.}, 2:16--81, 1992.

\bibitem{BoFo94}
A.~Borovkov and S.~Foss.
\newblock Two ergodicity criteria for stochastically recursive sequences.
\newblock {\em Acta Applicandae Mathematicae}, 34:125--134, 1994.

\bibitem{BoLa}
P.~Bougerol and J.~Lacroix.
\newblock {\em Products of Random Matrices with Applications to {S}chr\"odinger
  Operators}.
\newblock Progress in Probability and Statistics. Birk\"auser, 1985.

\bibitem{BFLi}
A.~Brandt, P.~Franken, and B.~Lisek.
\newblock {\em Stationary Stochastic Models}.
\newblock Prob. and Math. Stat. Wiley, New York, 1990.

\bibitem{BrVi}
M.~Brilman and J.M. Vincent.
\newblock Synchronisation by resources sharing : a performance analysis.
\newblock Technical report, MAI-IMAG, Grenoble, France, 1995.

\bibitem{CDQV83}
G.~Cohen, D.~Dubois, J.P. Quadrat, and M.~Viot.
\newblock Analyse du comportement p\'{e}riodique des syst\`{e}mes de production
  par la th\'{e}orie des dio\"\i des.
\newblock Technical Report 191, INRIA, 1983.

\bibitem{CDQV85}
G.~Cohen, D.~Dubois, J.P. Quadrat, and M.~Viot.
\newblock A linear system-theoretic view of discrete-event processes and its
  use for performance evaluation in manufacturing.
\newblock {\em IEEE Trans. Automatic Control}, AC-30:210--220, 1985.

\bibitem{cohe}
J.~Cohen.
\newblock Subadditivity, generalized product of random matrices and operations
  research.
\newblock {\em SIAM Review}, 30(1):69--86, 1988.

\bibitem{cuni62}
R.~Cuninghame-Green.
\newblock Describing industrial processes with interference and approximating
  their steady-state behaviour.
\newblock {\em Oper. Res. Quat.}, 13(1):95--100, 1962.

\bibitem{cuni79}
R.~Cuninghame-Green.
\newblock {\em Minimax Algebra}, volume 166 of {\em Lecture Notes in Economics
  and Mathematical Systems}.
\newblock Springer-Verlag, Berlin, 1979.

\bibitem{FuKe}
H.~Furstenberg and H.~Kesten.
\newblock Products of random matrices.
\newblock {\em Ann. Math. Statist.}, 31:457--469, 1960.

\bibitem{gaub94b}
S.~Gaubert.
\newblock On semigroups of matrices in the $(\max,+)$ algebra.
\newblock Technical Report 2172, INRIA, 1994.

\bibitem{GaMa95}
S.~Gaubert and J.~Mairesse.
\newblock Task resource models and (max,+) automata.
\newblock In J.~Gunawardena, editor, {\em Idempotency}. Cambridge University
  Press, 1995.

\bibitem{GlYa}
P.~Glasserman and D.~Yao.
\newblock {\em Monotone Structure in Discrete-Event Systems}.
\newblock John Wiley \& Sons, 1994.

\bibitem{GoMi77}
M.~Gondran and M.~Minoux.
\newblock Valeurs propres et vecteurs propres dans les dio\"\i des et leur
  interpr\'etation en th\'eorie des graphes.
\newblock {\em EDF, Bulletin de la Direction des Etudes et Recherches, Serie C,
  Math\'ematiques Informatique}, 2:25--41, 1977.

\bibitem{GoNe}
W.~Gordon and G.~Newell.
\newblock Closed queuing systems with exponential servers.
\newblock {\em Oper. Res.}, 15:254--265, 1967.

\bibitem{grif}
R.~Griffiths.
\newblock Frenkel-{K}ontorova models of commensurate-incommensurate phase
  transitions.
\newblock In H.~van Beijeren, editor, {\em Fundamental problems in statistical
  mechanics VII}. Elsevier Science Publishers, 1990.

\bibitem{KaMa92}
H.~Kaspi and A.~Mandelbaum.
\newblock Regenerative closed queueing networks.
\newblock {\em Stoch. and Stoch. Reports}, 39:239--258, 1992.

\bibitem{KaMa94}
H.~Kaspi and A.~Mandelbaum.
\newblock On {H}arris recurrence in continuous time.
\newblock {\em Math. Oper. Research (to appear)}, 1994.

\bibitem{loyn}
R.~Loynes.
\newblock The stability of a queue with non-independent interarrival and
  service times.
\newblock {\em Proc. Camb. Philos. Soc.}, 58:497--520, 1962.

\bibitem{mair95e}
J.~Mairesse.
\newblock {\em Stabilit\'e des syst\`emes \`a \'ev\'enements discrets
  stochastiques. Approche alg\'ebrique}.
\newblock PhD thesis, Ecole Polytechnique, Paris, 1995.
\newblock In english.

\bibitem{MeTw}
S.~Meyn and R.~Tweedie.
\newblock {\em Markov Chains and Stochastic Stability}.
\newblock Springer-Verlag, Berlin, 1993.

\bibitem{Oal90}
G.J. Olsder, J.~Resing, R.~de~Vries, M.~Keane, and G.~Hooghiemstra.
\newblock Discrete event systems with stochastic processing times.
\newblock {\em IEEE Trans. on Automatic Control}, 35(3):299--302, 1990.

\bibitem{Ral90}
J.~Resing, R.~de~Vries, G.~Hooghiemstra, M.~Keane, and G.J. Olsder.
\newblock Asymptotic behavior of random discrete event systems.
\newblock {\em Stoch. Proc. and Applications}, 36:195--216, 1990.

\bibitem{roma}
I.V. Romanovski\u\i.
\newblock Optimization and stationary control of discrete deterministic process
  in dynamic programming.
\newblock {\em Cybernetics}, 3:66--78, 1967.

\bibitem{YaKo}
S.~Yakovenko and L.~Kontorer.
\newblock Nonlinear semigroups and infinite horizon optimization.
\newblock In V.~Maslov and S.~Samborski\u\i, editors, {\em Idempotent
  analysis}, volume~13 of {\em Adv. in Sov. Math.} AMS, 1992.

\end{thebibliography}
%% \bibliographystyle{plain}

\end{document}